\documentclass[conference]{IEEEtran}
\IEEEoverridecommandlockouts
\usepackage{cite}
\usepackage{amsmath,amssymb,amsfonts}
\usepackage{algorithmic}
\usepackage{graphicx}
\usepackage{textcomp}
\usepackage{xcolor}
\usepackage{amsmath,amsfonts}
\usepackage{algorithmic}
\usepackage{array}
\usepackage{textcomp}
\usepackage{stfloats}
\usepackage{url}
\usepackage{verbatim}
\usepackage{graphicx}
\usepackage{cite}
\usepackage[T1]{fontenc}
\usepackage{tabularx}
\usepackage{graphicx}
\usepackage[cmintegrals]{newtxmath}
\usepackage{algorithmic}
\usepackage{caption}
\usepackage{cite}
\usepackage{setspace} 
\usepackage{subcaption}
\usepackage{xcolor}
\usepackage[para]{footmisc}
\usepackage{tabularx,array}
\usepackage{multirow}
\usepackage{booktabs}
\usepackage[linesnumbered,ruled,vlined]{algorithm2e}
\SetKwInput{KwInput}{Input}                
\SetKwInput{KwOutput}{Output}  
\SetKwInput{KwFunction}{Function}  
\SetKwInput{KwReturn}{Return}  
\usepackage{textcomp}
\usepackage{linguex}
\usepackage[linguistics,edges]{forest}
\usepackage{metalogo}
\usepackage{url}
\usepackage{csquotes} 
\usepackage{arydshln} 
\usepackage{color, soul}
\usepackage{amssymb}
\usepackage{makecell}
\usepackage{etoolbox}
\usepackage{threeparttable}
\usepackage{pifont}
\usepackage{float}
\usepackage[normalem]{ulem}
\usepackage{mathtools}

\renewcommand{\IEEEbibitemsep}{0pt plus 0.5pt}
\makeatletter
\IEEEtriggercmd{\reset@font\normalfont\fontsize{7pt}{8.40pt}\selectfont}
\makeatother
\IEEEtriggeratref{1}
\usepackage{comment}
\usepackage{balance}

\begin{document}

\title{Optimizing Vehicle-to-Edge Mapping with Load Balancing for Attack-Resilience in IoV }
\author{\IEEEauthorblockN{Anum~Talpur~and~Mohan~Gurusamy}
\IEEEauthorblockA{\textit{Department of Electrical and Computer Engineering,} \\
\textit{National University of Singapore, Singapore}\\
Email: anum.talpur@u.nus.edu, gmohan@nus.edu.sg}}

\maketitle

\begin{abstract}
Attack-resilience is essential to maintain continuous service availability in Internet of Vehicles (IoV) where critical tasks are carried out. In this paper, we address the problem of service outage due to attacks on the edge network and propose an attack-resilient mapping of vehicles to edge nodes that host different types of service instances considering resource efficiency and delay. The distribution of service requests (of an attack-affected edge node) to multiple attack-free edge nodes is performed with an optimal vehicle-to-edge (V2E) mapping. The optimal mapping aims to improve the user experience with minimal delay while considering fair usage of edge capacities and balanced load upon a failure over different edge nodes. The proposed mapping solution is used within a deep reinforcement learning (DRL) based framework to effectively deal with the dynamism in service requests and vehicle mobility. We demonstrate the effectiveness of the proposed mapping approach through extensive simulation results using real-world vehicle mobility datasets from three cities. 
\end{abstract}


\section{Introduction}
Edge network (EN) is an emerging innovation for 5G-enabled IoV networks to alleviate the problem related to latency and reliability by bringing the computation capabilities and storage near to the end-users. Although the EN paradigm addresses these problems, it introduces security risks where edge nodes are exposed to different outage attacks such as jamming attack and denial of service attack \cite{edgeattacks}. Such attacks cause an edge node to fail and pose serious threats since a malicious edge node can compromise the service availability to vehicles where vehicles are performing the driving tasks and service disruption may result in catastrophe and danger to human lives. An important requirement is to ensure attack resilience with the ability to withstand node failures and maintain a good quality of service in the system. \par 
Several studies exist on defense against outage attacks \cite{useEdge1,useEdge2}, but the resilience against edge node failure is less explored. Some studies have approached the edge attack problem from the perspective of resilience with different objectives and they do not consider the impacts of dynamic and mobile traffic of IoVs \cite{gametheoryDOSdetection,GaussianDoSedgeAttack}. The use of the backup resource (BR) reservation method is well-known in the literature to handle server failures with resilience \cite{backup3,backup5}. However, reservation of resources is expensive and results in wastage, and it is not desirable especially when we are dealing with the limited resource EN. There are some recent works where different approaches are used to minimize the backup cost \cite{backup4,backup6}, but the drawbacks of backup reservation still persist. Different from the above works, we propose pro-active attack-resilient V2E mapping to ensure minimal or no disruption of service availability without using backup resource reservation. \par 
\begin{figure}[htbp]
	\centering
	\includegraphics[width=2.2in, height=1.8in]{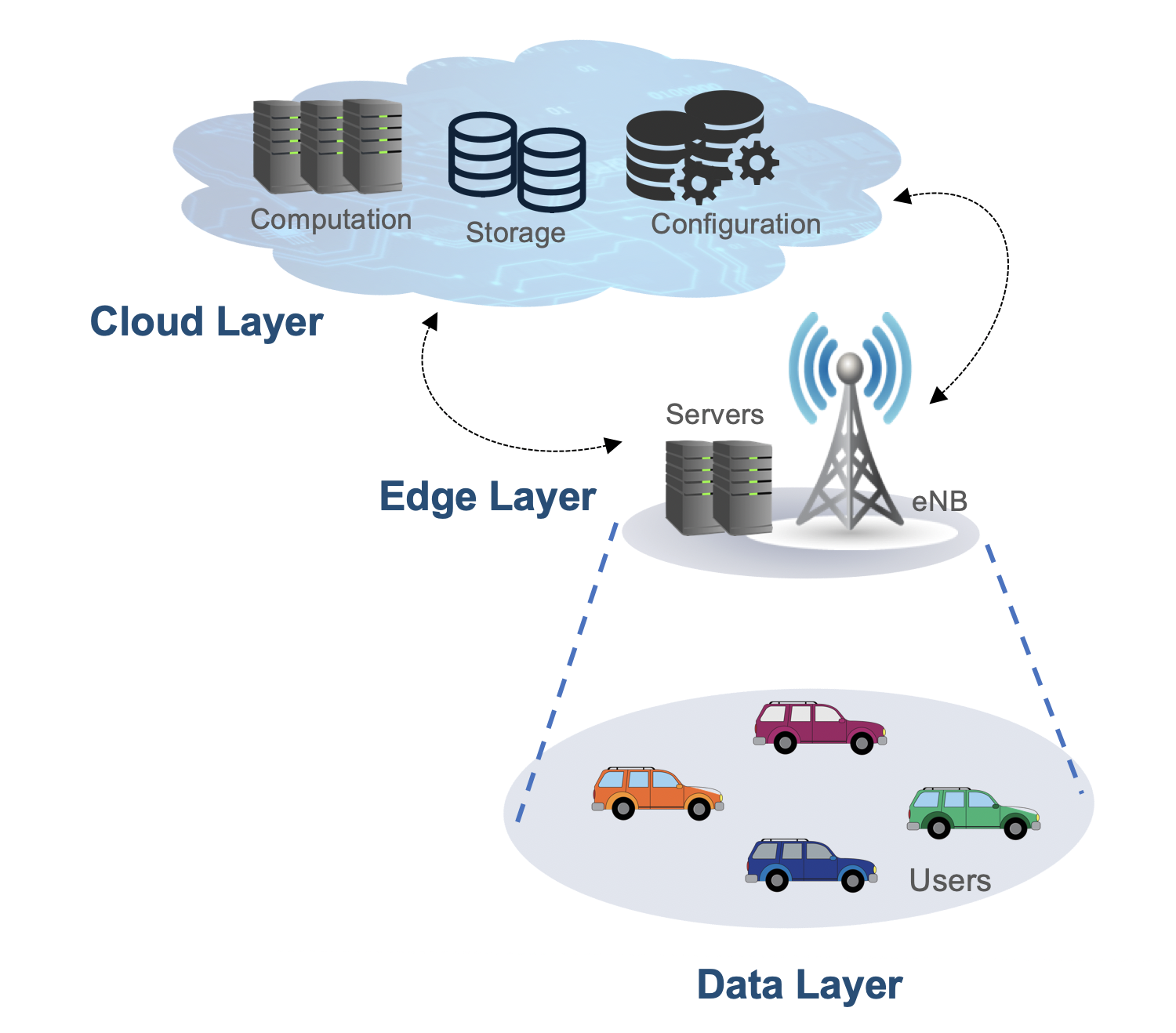}
	\caption{Architecture of an edge-enabled IoV network}
	\label{fig:Architecture}
\end{figure}
In this paper, we consider an edge-based IoV network where different types of services are deployed over EN to assist vehicles in different applications, as shown in Fig. \ref{fig:Architecture}. The outage attack over the edge node causes all service instances (SIs) running on that node to fail. The mapping of a vehicle to an edge node hosting the required service instance during the normal working condition is termed as primary V2E mapping, whereas the mapping upon a node failure is termed as secondary V2E mapping. To ensure service availability upon a node failure, we develop secondary V2E mapping formulations that fulfil the attack-resilient conditions in the design framework. We start with the service placement to find the optimal placement of services at the edge servers and calculate the primary vehicle-to-edge (V2E) mappings used in normal working conditions. The primary mappings tend to minimize delay observed by the vehicles while accessing the service. Together with primary mappings, another set of mappings (i.e. secondary V2E mappings) is computed in a proactive manner. The secondary mappings are used upon a node failure to ensure the service availability to attack-affected vehicles. The secondary mappings are temporary mappings and chosen a-priori to maintain disruption-free service availability by using the resources of the currently working servers (not affected by the attack) to provide best possible service quality within the constraints of available resources. This will be followed by a recovery phase where new servers (if needed) and updated service placement may be employed in the post-attack scenario. \par 
In this paper, we adapt our existing resilient service placement (RSP) framework presented in \cite{AnumRSP} and develop new secondary mapping optimization formulations to enhance the network performance when the attack is active and recovery is not completed. We thus propose a load-balanced proactive secondary V2E mapping (LB-PSVM) model in this paper that maps attack-affected vehicles to attack-free edge servers while satisfying minimum service delay and distributing the affected service instance loads in a balanced way among different edge nodes in terms of the usage of processing capabilities. We develop programming formulation for LB-PSVM and use within a DRL based framework to effectively deal with the vehicle's mobility and dynamics in service requests using real-world vehicle trajectories from three cities. Finally, we carry out a performance study on multiple datasets to analyze the impacts of attack-resilient LB-PSVM on service delay and edge processing capacity.

\section{System Description and Problem Statement}

\subsection{System Model}
\label{Sec:SystemModel}
We consider a three-layer architecture of the edge-enabled IoV network, as shown in Fig. \ref{fig:Architecture}. First, there is a \textit{\textbf{data layer}} where we have a city road environment with a real journey of vehicles that connect with the edge layer referred to as "V2E mapping" and avail different types of services to carry out driving tasks. Vehicles generate a service request as a 4-tuple structure $<v\,,\, l\,,\, t\,,\, s>$ where $t$ is the time at which the request is generated and $l$ is the location of the vehicle $v$ requesting for service type $s$. We use real-world datasets in this paper \cite{sanfrancisco,beijing1,rome} in which the vehicles are mobile and equipped with necessary sensors to provide relevant information. Each service type has a stringent delay $D_s$ and resource requirements $R_s$ needed to instantiate a service. The number of vehicles requesting service $s$ is uniformly distributed (denoted as $\lambda_s$) and $I_s$ denotes the number of instances required for service type $s$. The second layer is the \textit{\textbf{edge layer}} using evolved NodeB (eNB) stations forming a multi-cell coverage area for mobile vehicles. Each edge node is comprised of edge servers with an abstract measure of resources $C_e$, used to run different services. Each edge node can deploy multiple service instances (SIs) with the processing capacity $\mathbb{C}$ for each instance. The attacks are assumed to take place at the edge layer. Additionally, the edge layer connects to the large capacity \textit{\textbf{cloud layer}} to download and perform service instantiation. We assume adequate links are available to enable communication among different layers.\par 

\subsection{Proposed Resilient Service Placement (RSP) Framework}
\label{Sec:resilientplacement}
\begin{figure}[htbp]
	\centering
	\includegraphics[width=2.6in, height=1.5in]{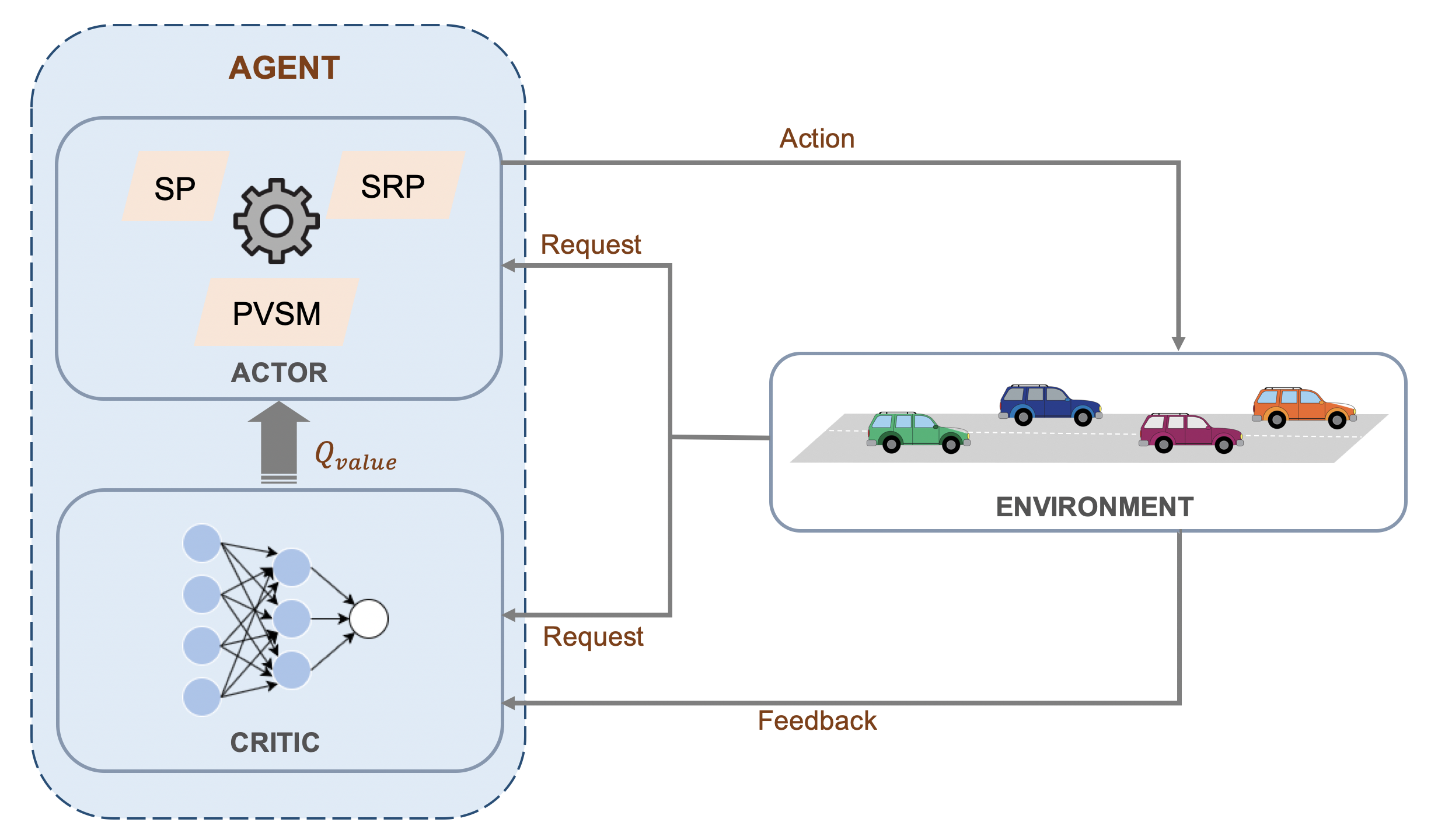}
	\caption{Resilient service placement framework}
	\label{fig:framework}
\end{figure}

Based on the system model above, we will explain our RSP framework from our earlier work in \cite{AnumRSP}, which is adapted and extended in this paper. The key idea is to perform an attack-resilient optimal service placement to ensure disruption-free service availability to the vehicles. The architecture of our RSP framework is shown in Fig. \ref{fig:framework}. It exploits an actor-critic DRL model along with optimization formulations. The DRL agent learns and updates the actor-critic network by interacting with the dynamic and time-varying IoV environment. An actor is a primary function which generates actions, and a critic estimates a quality value ($Q_{value}$) needed to keep the performance of an actor updated. The functioning of the actor and critic network is further explained below. \par 

\subsubsection{Actor Network}
\label{Sec:actor}
In our actor design, we leverage the optimization formulations to perform optimal service placements, V2E mappings, and recovery placements in a dynamic manner. We use three programming formulations for three different but related problems. 
\begin{itemize}
	\item \textbf{\textit{Problem 1 (Service Placement (SP)):}} The SP deals with the problem of deciding on optimal edge locations $x_e^s$ to instantiate service instances of different requested service types, subject to resilience, delay and resource constraints, and considering the vehicle's mobility and dynamics in the service requests.
	\item \textbf{\textit{Problem 2 (Proactive Secondary V2E Mapping (PSVM))}} Once the service placements are completed, a vehicle receives its requested service from a server based on primary V2E mapping subject to lower service delay. Upon an attack, it needs to get service from an attack-free server chosen as a secondary V2E mapping node. Therefore, given a set of optimal edge locations $x_e^s$ obtained from SP, the PSVM problem is to find the secondary mappings subject to minimal delay and ensure continuous service availability for the set of vehicles impacted by the attack. 
	\item \textbf{\textit{Problem 3 (Service Recovery Placement (SRP)):}} The SRP deals with the recovery of attacked SIs to the attack-free edge nodes. Given a set of optimal edge locations $x_e^s$, the problem is to find optimal recovery placement locations of SIs of each service type with their resource and delay requirements such that the recovery placements are different from $x_e^s$ and use minimal edge resources. While PSVM uses the resources of unaffected existing SIs hosted on the working servers, SRP deals with re-instantiating the affected SIs on the existing or new servers.
\end{itemize}
We describe the functioning of the agent network that uses the above three formulations for different network states. For the normal working \textbf{\textit{attack-free}} state, the SP deals with the primary placement of services. One of the important constraints for the SP solution is the redundancy to ensure the availability of the same type of service from multiple edge nodes. Along with SP, we solve PSVM in a proactive manner but the mappings are used only when the network is under attack. It helps to maintain resilience and avoid service breakage for vehicles. In the state of \textbf{\textit{attack}}, our framework promptly maps attack-affected vehicles to the existing unaffected service instances on the attack-free servers using PSVM to ensure service availability. In the post-attack scenario, it solves the SRP model to find the optimal instance recovery placements to re-instantiate the affected service instances on the attack-free server, in the given state of the environment. Once the SRP is solved, the attacked SIs are re-instantiated and the network becomes resilient to attack with a negligible loss in the network performance. 

\subsubsection{Critic Network}
In our design, the critic network is a neural network (NN) model to estimate the network quality based on the feedback received from the environment. The feedback is a response to the action taken by an actor network. The critic network updates its parameters to minimize the mean square loss function based on the feedback and network state, and estimates a $Q_{value}$ which is used to maintain good performance of the network by dynamically re-optimizing the actor problem formulations. The $Q_{value}$ changes in the range 0-1 and the lower the value is, the poorer the system performance, necessitating re-optimizations.

\subsection{Problem Statement} 
\begin{figure}[htbp]
	\centering
	\includegraphics[width=2.6in, height=1.7in]{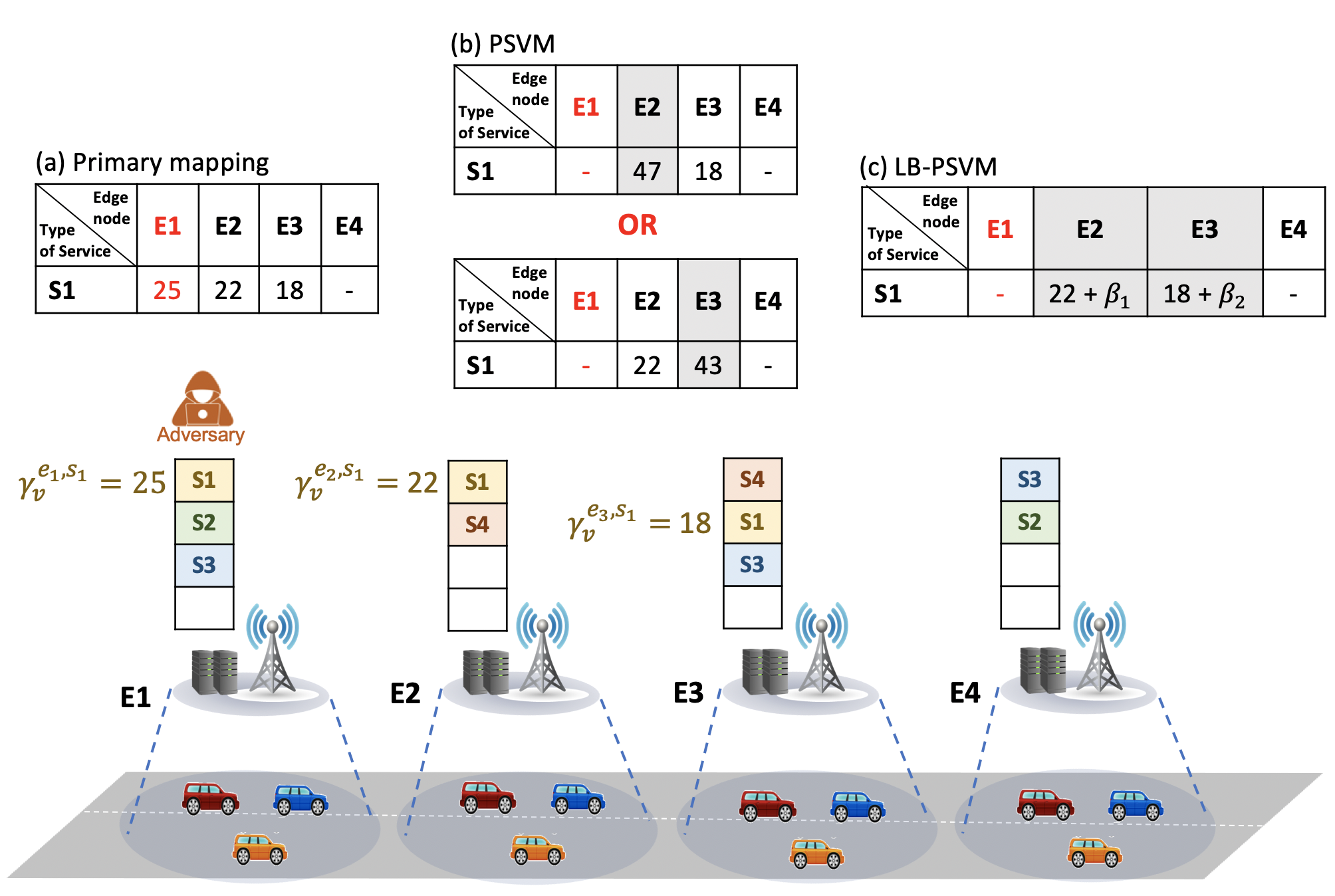}
	\caption{Secondary V2E mapping example with E1 under attack}
	\label{fig:example}
\end{figure}

We start with an illustrating example in Fig. \ref{fig:example}. We consider a network of 4 edge nodes (i.e. $E1$, $E2$, $E3$, and $E4$) placing instances of 4 different service types (i.e. $S1$, $S2$, $S3$, and $S4$). The primary mapping for service $s$ at edge node $e$ is represented as $\gamma_v^{e,s}$. Assume that node $E1$ is under attack where an instance of service $S1$, $S2$, and $S3$ is deployed, the problem is to remap the attack-affected traffic for all the three service types to attack-free nodes. \par 
We consider the scenario wherein the number of vehicles availing service $S1$ at $E1$ is 25, at $E2$ is 22 and at $E3$ is 18. Given the $E1$ attack scenario, we need to remap 25 vehicles and communicate the next suitable $S1$ instance locations to the vehicles. In PSVM, the decision on a secondary server is subject to minimal delay and remaps all of the attacked-affected vehicles to the lowest delay server. As shown in Fig. \ref{fig:example}, when $E2$ is the lowest delay secondary server, the traffic distribution among attack-free instances is 47 and 18. On the contrary, for $E3$ as a secondary server, the traffic distribution is 22 and 43. For the simplicity purpose, PSVM only considers the delay while deciding on mapping locations and ignores the edge performance which results in inefficient usage of processing capacities. We thus propose LB-PSVM that remaps attack-affected traffic to more than one servers and balances the load among edge nodes along with minimal service delay. The balanced mapping of vehicles among attack-free servers will help prevent congestion at single server and avoids the creation of long queues causing them to experience higher delays. In the view of above example, the problem statement for LB-PSVM is to calculate $\beta_i$ where,
\begin{equation}
\beta_i \ge 0; \ \ \ \    \forall i \epsilon n
\label{eqbeta}
\end{equation}
The expected output is to decide $\beta_i$ sets of V2E mappings where $i=1,2,...n$ and $n$ represents the number of available attack-free edge servers/nodes.

\section{Proposed Optimization Formulation for V2E Mapping}
\label{Sec:V2Emapping}
Given the above problem statement, we present the optimization formulation for LB-PSVM. We start with the computation of the primary V2E mappings used during the normal operation. The primary V2E mappings at time unit $t$ are used as an input for the LB-PSVM to calculate proactive secondary mappings for the attack that may take place at time unit $t+1$. The primary V2E mappings aim to minimize the maximum delay for vehicles following the optimal location $x_e^s$ of SIs we have from Problem 1 (of Section \ref{Sec:actor}). We use the real variable $\gamma_v^{e,s}$ to express the number of vehicles primarily mapped to different SIs at different edge nodes during normal working conditions. Then, the objective is formulated as,
\begin{equation}
\underset{\gamma}{minmax} \ \left(\sum_{e\in E} \sum_{s\in S} d_e^s\right) \gamma_v^{e,s}
\label{eq:primaryV2E}
\end{equation}
Here, $d_{e}^s$ is the average service delay calculated as a propagation delay observed by vehicles while requesting service $s$ from edge node $e$. The calculation of primary V2E mappings is subject to multiple constraints. \par 
\begin{equation}
\sum_{e\in E}\gamma_v^{e,s} = \lambda_s, \   \ \forall s\in S
\label{EQ:primaryC1}
\end{equation}
\begin{equation}
\gamma_v^{e,s} \le \mathbb{C} x_e^s, \  \forall s\in S, \forall e\in E
\label{EQ:primaryC2}
\end{equation}
\begin{equation}
\gamma_v^{e,s} \ge 0, \  \forall s\in S, \forall e\in E
\label{EQ:primaryC3}
\end{equation}
Constraint (\ref{EQ:primaryC1}) guarantees the sum of the set of the vehicles mapped to different SIs at different edge nodes must be equal to the corresponding arrival requests. Constraint (\ref{EQ:primaryC2}) ensures the vehicles must be mapped to the edge nodes where their requested service is placed and the number of vehicles mapped must not exceed the processing capacity $\mathbb{C}$. Finally, condition (\ref{EQ:primaryC3}) defines the decision variable $\gamma_v^{e,s} $ as a real variable. \par
Next, using the primary mapping calculated from (\ref{eq:primaryV2E}), we aim to optimally and proactively calculate the secondary V2E mappings (i.e. LB-PSVM) which will be used when the network is under attack to maintain resilient service provisioning for vehicles. The formulation of LB-PSVM consists of two parts: (i) to maximize the fair and balanced distribution of processing load among different edge nodes while performing V2E mappings, and (ii) to minimize the delay experienced by vehicles while being mapped to attack-free edge nodes. We start with the first part where the goal is to construct a mapping of vehicles to each SI in a way that the request processing load at different edge nodes is fairly balanced. We use a well-known logarithmic utility for load balancing, also known as proportional fairness \cite{ProportionFairness}. In our context, it is defined as the weighted sum of the logarithm of the number of vehicles mapped.
Assume that $\beta_i$ represents the secondary mapping of $\beta$ number of vehicles at $i^{th}$ edge node. The first part of our objective for LB-PSVM is formulated as,
\begin{equation}
max \ f_1(\beta)
\label{EQ:secondaryP}
\end{equation}
where,
\begin{equation}
f_1(\beta) = \sum_{i=1}^{n} w_i \log \beta_i
\label{EQ:secondaryP1}
\end{equation}
Here, $n=I_s - 1$ and $w_i$ is the weight that represents the available processing capacity of the edge node. Therefore, in our formulation, $w_i$ is calculated as,
\begin{equation}
w_i = 1 - \frac{\gamma_v^{e_i,s}}{\mathbb{C}}
\label{EQ:weight}
\end{equation}
Here, $\gamma_v^{e_i,s}$ represents the primary mappings at the edge node $e_i$ for service type $s$ and $\mathbb{C}$ is the processing capacity of the edge node. Thus, $f_1(\beta)$ is rewritten as,
\begin{equation}
f_1(\beta) = \sum_{i=1}^{n} \left(1 - \frac{\gamma_v^{e_i,s}-\varepsilon}{\mathbb{C}}\right) \log \beta_i, \  \forall s\in S
\label{EQ:secondaryP1-2}
\end{equation}
We use a very small offset $\varepsilon$ to avoid unnecessary situations where the number of primarily mappings $\gamma_v^{e_i,s}$ is equal to $\mathbb{C}$ which would result in multiplication by zero in the objective function. The constraints associated with the above optimization objective are,
\begin{equation}
\sum_{i=1}^{n} \beta_i = \grave{\gamma}_v^{e,s}, \  \forall s\in S
\label{EQ:SecondaryC1}
\end{equation}
\begin{equation}
\beta_i \ge 0, \  \forall i\in n
\label{EQ:SecondaryC2}
\end{equation}
Here, $\grave{\gamma}_v^{e,s}$ is the total number of vehicles primarily mapped to the attacked node for service type $s$. Hence, constraint (\ref{EQ:SecondaryC1}) ensures the secondary mapping of the number of vehicles must be equal to the number of vehicles affected by the attack. Constraint (\ref{EQ:SecondaryC2}) defines the domain of decision $\beta_i$ as a real variable. In the second part of LB-PSVM formulation, we aim to minimize the service delay experienced by attack-affected vehicles. The delay minimization objective considers two types of delays. First, we use propagation delay to consider the impact of changing mobility in our optimization model.
\begin{equation}
f_2(\beta) = \sum_{i=1}^{n} d_{e_i}^s \beta_i, \  \forall s\in S
\label{EQ:delaypropSec}
\end{equation}
Second, we use queuing delay to consider the network state where the arrival of requests is more than the processing capacity of the edge server. In such cases, our edge computation model works as an M/D/1 queue where arrival is according to a Markov stochastic model. The service processing rate is deterministic which is $\mathbb{C}$. We assume no queues when the arrival is less than $\mathbb{C}$. On the contrary, a queue will be created if the arrival is greater than $\mathbb{C}$. Assume $\lambda_s$ is the arrival rate, then the waiting time in a queue over edge server is calculated as \cite{queueMD1},
\begin{equation}
d_{queue}= 
\begin{cases}
\frac{\grave{\lambda_s}}{2\mathbb{C}(\mathbb{C}-\grave{\lambda_s})},& \text{if } \lambda_s > \mathbb{C}\\
0,              & \text{otherwise}
\end{cases}
\label{EQ:queue}
\end{equation}
Here, $\grave{\lambda_s}$ represents the number of vehicles in the queue and calculated as $\lambda_s-\mathbb{C}$. In our model, the total arrival at the attack-free edge node is equal to the sum of existing primary mappings and secondary V2E mappings, and calculated as $\lambda_s = \gamma_v^{e_i,s}+\beta_i$. Finally, the queuing delay is,
\begin{equation}
f_3(\beta) = \sum_{i=1}^{n}  \frac{\gamma_v^{e_i,s}+\beta_i-\mathbb{C}}{2\mathbb{C}(\mathbb{C}-(\gamma_v^{e_i,s}+\beta_i-\mathbb{C}))}, \  \forall s\in S
\label{EQ:delayqueueSec}
\end{equation}
This makes our second objective as,
\begin{equation}
min \ f_2(\beta)+f_3(\beta) 
\label{EQ:Objpart2}
\end{equation}
\hspace*{3mm} subject to,
\begin{equation}
\ f_2(\beta)+f_3(\beta) \le D_s
\label{EQ:SecondaryC3}
\end{equation}
where,
\begin{equation}
\ f_2(\beta)+f_3(\beta) = \sum_{i=1}^{n} d_{e_i}^s \beta_i + \frac{\gamma_v^{e_i,s}+\beta_i-\mathbb{C}}{2\mathbb{C}(\mathbb{C}-(\gamma_v^{e_i,s}+\beta_i-\mathbb{C}))}, \  \forall s\in S
\label{EQ:Objpart2-}
\end{equation}
The constraint in (\ref{EQ:SecondaryC3}) ensures that the delay experienced by attack-affected vehicles during secondary mapping should be less than the maximum allowable delay threshold $D_s$. Finally, we summarize the complete optimization programming formulation of the problem of optimal LB-PSVM model with maximal balancing of edge load and minimal service delay as,
\begin{equation}
Minimize: \sum_{i=1}^{n} -f_1(\beta_i) + k_1f_2(\beta_i) + k_2f_3(\beta_i)
\label{EQ:final}
\end{equation}
\hspace*{6mm} subject to: (\ref{EQ:SecondaryC1}), (\ref{EQ:SecondaryC2}) and (\ref{EQ:SecondaryC3}). \\ \\
Here, $k_1$ and $k_2$ are the scaling constants and $n$ is the number of attack-free edge nodes.

\section{Performance Study}
\subsection{Experimental settings}
\begin{figure}[hbt!]
	\captionsetup[subfigure]{justification=centering}
	\centering
	\begin{subfigure}{.22\textwidth}
		\centering
		\includegraphics[width=1.6in,height=1.2in]{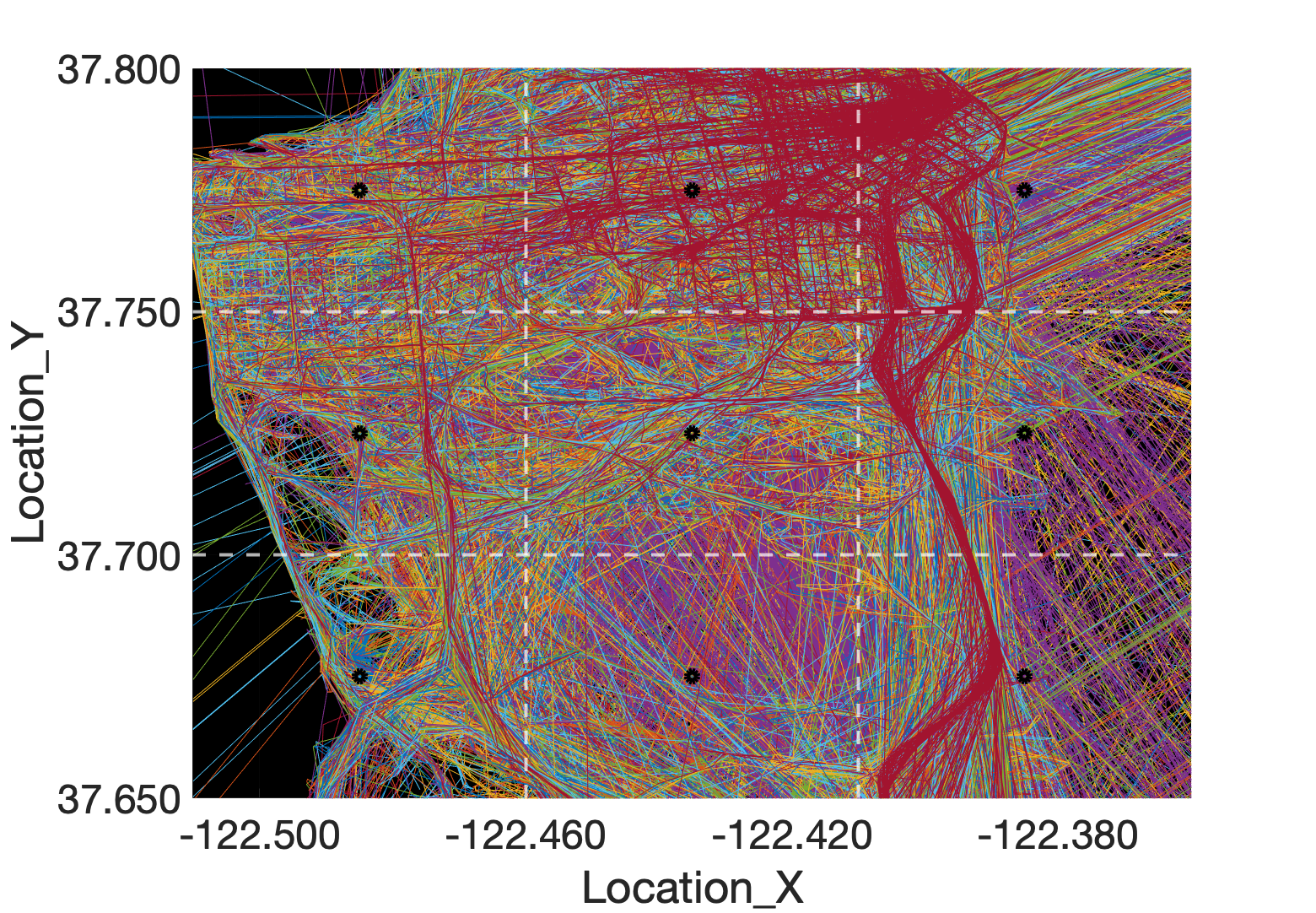}  
		\caption{San Francisco}
		\label{fig:data1}
	\end{subfigure}
	\begin{subfigure}{.22\textwidth}
		\centering
		\includegraphics[width=1.6in,height=1.2in]{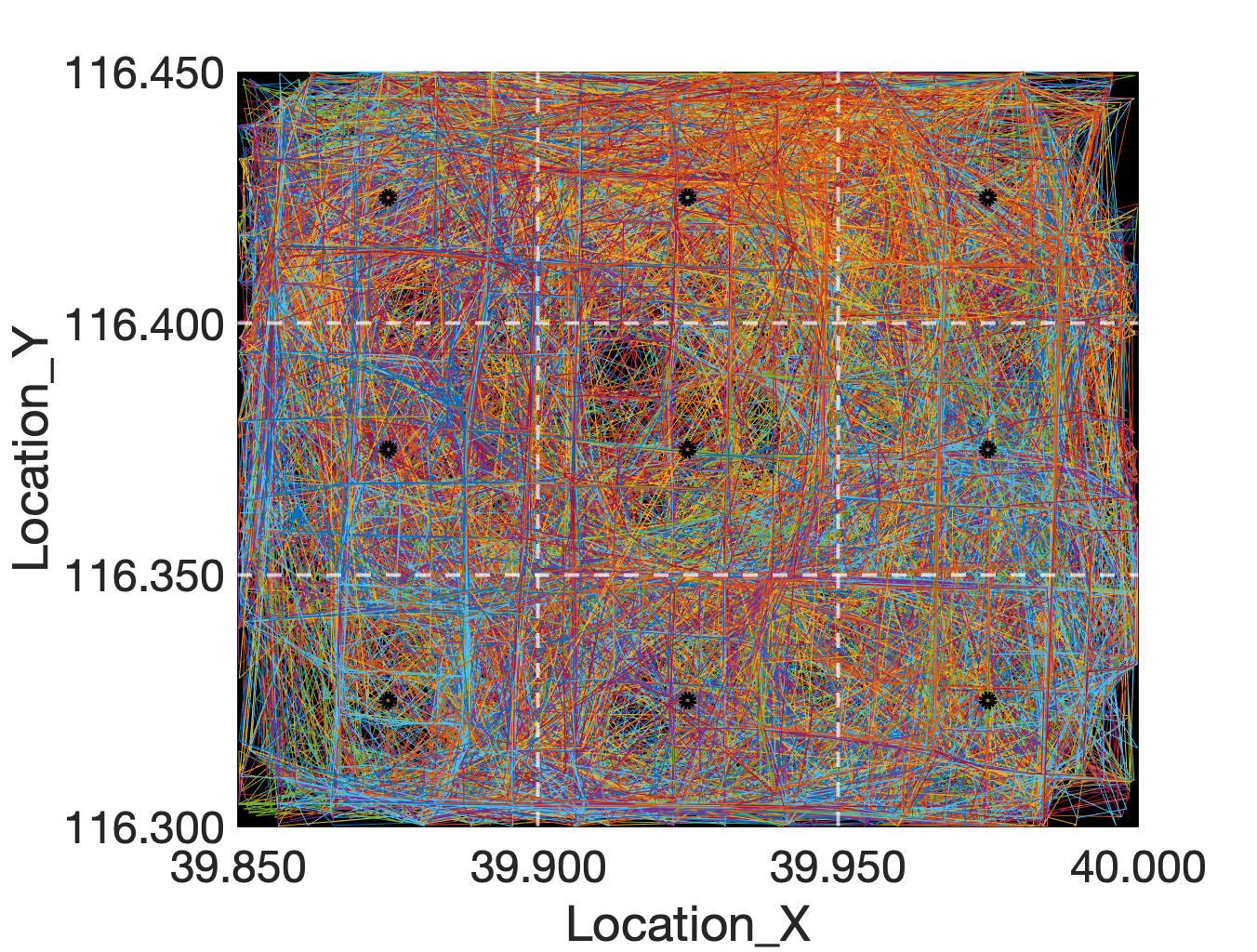}  
		\caption{Beijing}
		\label{fig:data2}
	\end{subfigure} 
	\begin{subfigure}{.22\textwidth}
		\centering
		\includegraphics[width=1.6in,height=1.2in]{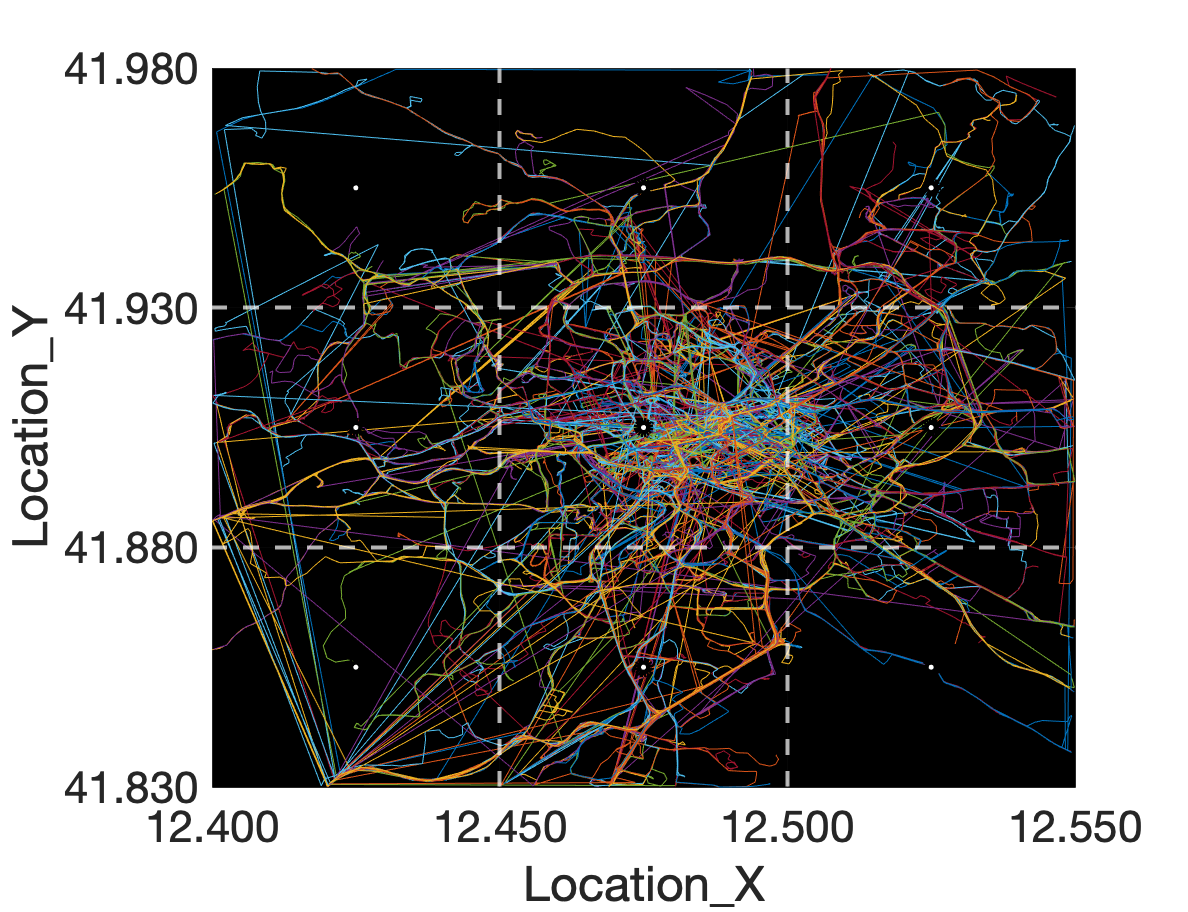}  
		\caption{Rome}
		\label{fig:data3}
	\end{subfigure}
	\caption{Vehicle trajectories for different city environments}
	\label{fig:dataset}
\end{figure}

The implementation of our proposed framework is carried out in MATLAB using 3 real-world vehicle mobility datasets for the city of San Franciso \cite{sanfrancisco}, Rome \cite{rome}, and Beijing \cite{beijing1}. We use multiple datasets to validate the effectiveness of our model for dynamic traffic conditions. The trajectories are generated from a maximum of 500, 500, and 194 taxis in the San Francisco, Beijing, and Rome, respectively. From the given datasets, we extract an area of same size (i.e. 15x15 $km^2$) from each dataset for a fair comparison. We consider an edge system with 9 areas as shown in Fig. \ref{fig:dataset}, where each area has an edge node $e \in E$ with an abstract measure of resources $C_e$ = 100 units. The edge network provides 8 different types of services and the placement of a single instance of service type $s$ requires $R_s = {10, 12, 14, ...24}$ resource units, and has delay threshold of $D_s = {50, 60, 70, ..., 120}$, respectively. The processing capacity of each service instance to provide simultaneous connections to vehicles is $\mathbb{C} = 30$. \par 
We test and run LB-PSVM in the DRL-based resilient service placement framework, defined in Section \ref{Sec:resilientplacement}. For DRL, the critic NN is a feed-forward network with 4 hidden layers, each with 512, 256, 64, and 32 neurons respectively. To avoid overfitting, the learning rate of 0.01 is used and, the maximum number of episodes is 1500 with each episode having a maximum of 20 iterations and a batch size of 100 to train a network. The $Q_{value}$ by the critic is examined every 5 time units. In case of poor performance, the programming models of the actor are re-optimized. All experiments are evaluated on an Intel Core i5 2GHz and 8GB RAM system.

\subsection{Comparison}
\label{Sec:comparison}
We compare the performance of LB-PSVM against two approaches. The first approach is the PSVM model, where all of the attack-affected vehicles are mapped to single optimal attack-free server, as discussed above. The second approach is a well-known resilience technique of backup resource (BR) reservation \cite{backup3, backup4,backup5}. In the BR, we install and reserve one extra instance of each service type for the attack-resilience. Note that the backup resources remain idle until failure occurs.

\subsection{Results}
In this section, we evaluate and validate the LB-PSVM against SP, SRP and PSVM, using difference performance metric where attack takes place at every $100^{th}$ time unit. Note that the SP is used in a pre-attack and SRP, PSVM and LB-PSVM are for the post attack scenarios. \par 
\begin{figure}[hbt!]
	\captionsetup[subfigure]{justification=centering}
	\centering
	\begin{subfigure}{.22\textwidth}
		\centering
		\includegraphics[width=1.55in,height=1.15in]{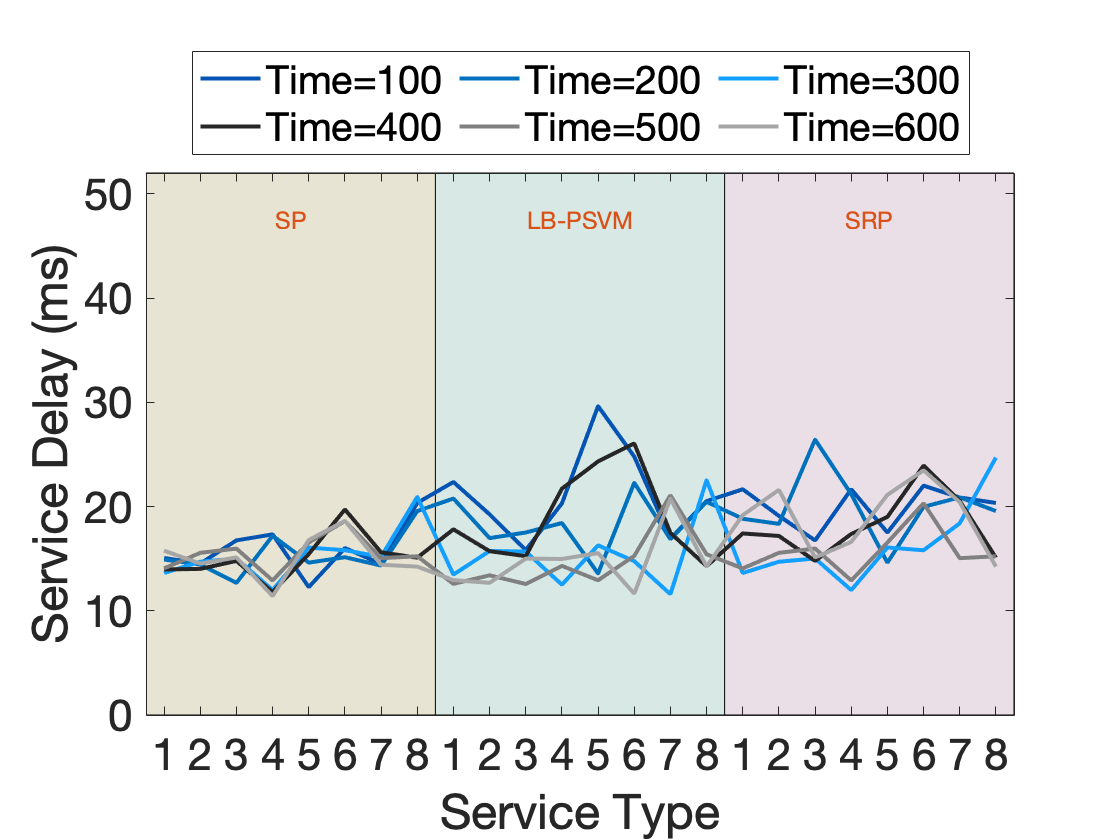}  
		\caption{San Francisco}
		\label{fig:Delay-SF}
	\end{subfigure}
	\begin{subfigure}{.22\textwidth}
		\centering
		\includegraphics[width=1.55in,height=1.15in]{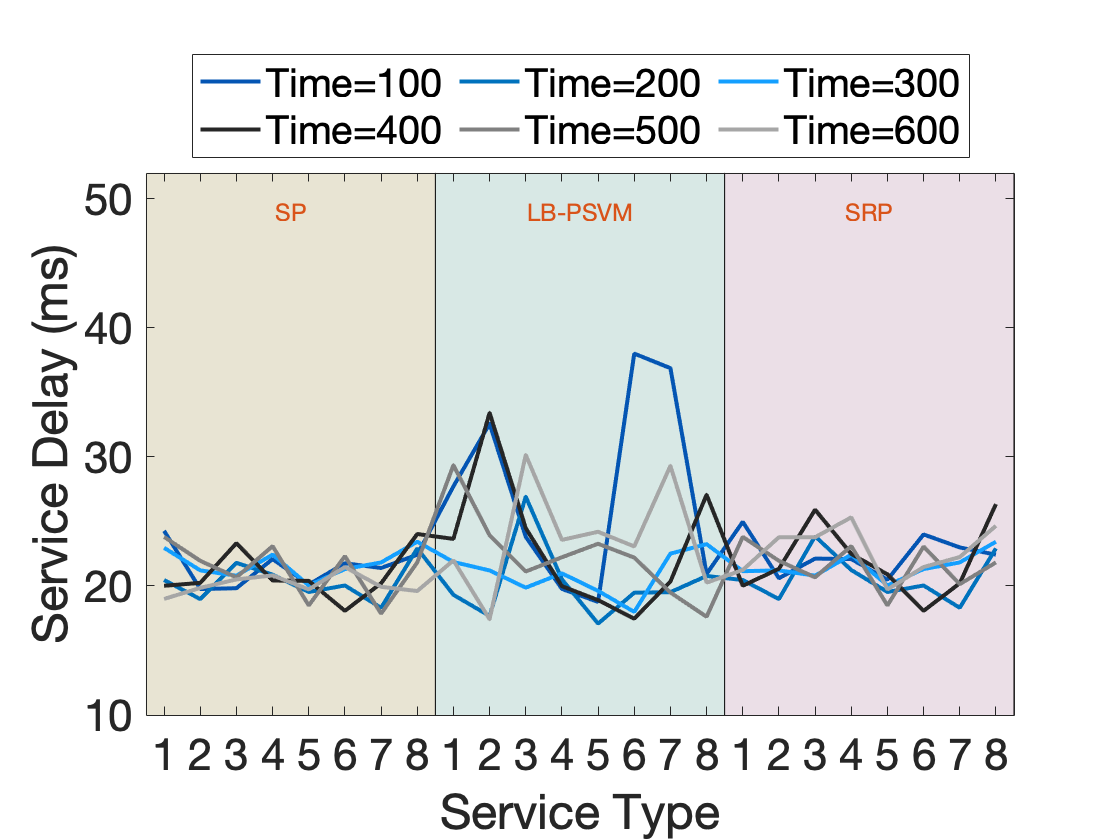}  
		\caption{Beijing}
		\label{fig:Delay-BJ}
	\end{subfigure} \\
	\begin{subfigure}{.22\textwidth}
		\centering
		\includegraphics[width=1.55in,height=1.15in]{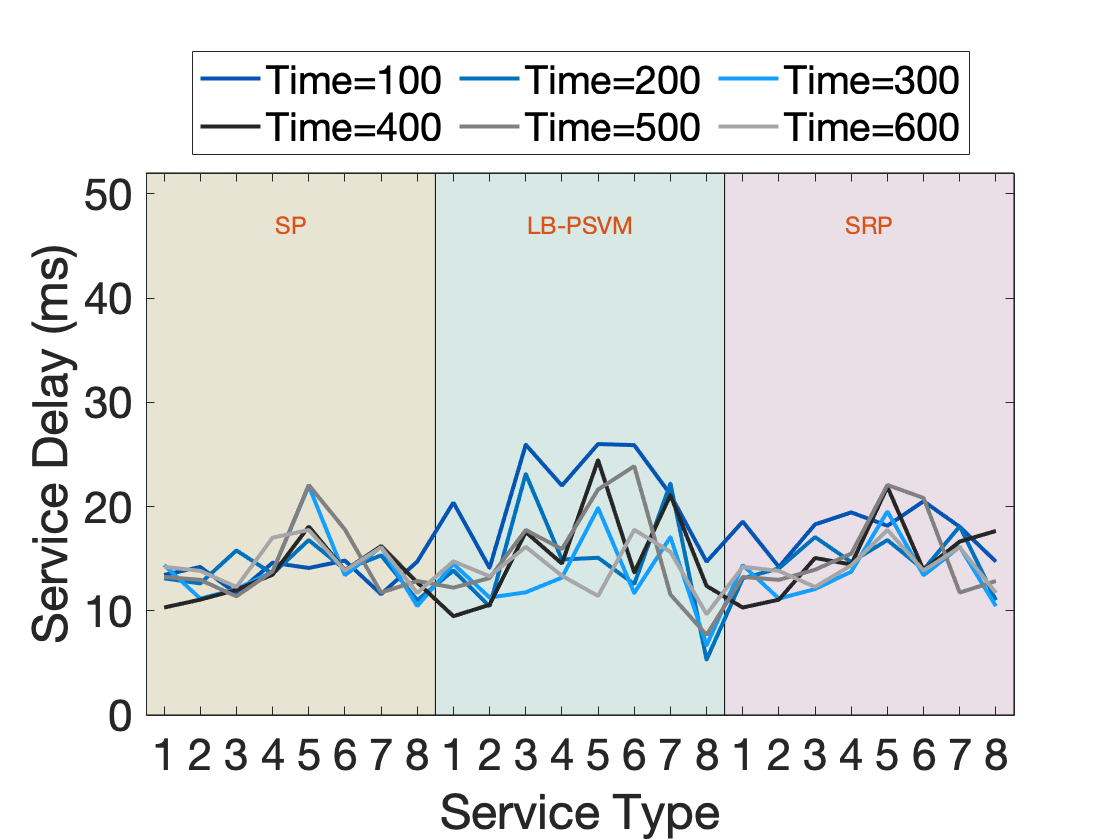}  
		\caption{Rome}
		\label{fig:Delay-RM}
	\end{subfigure}
	\caption{Service Delay (ms)}
	\label{fig:Delay}
\end{figure}

Fig. \ref{fig:Delay} investigates the service delay for LB-PSVM, validating its performance against SP and SRP. For the city of San Francisco, it can be observed from Fig. \ref{fig:Delay-SF} that the delay is relatively lower in SP compared to post-attack network conditions. Once the attack is launched, our framework activates LB-PSVM to maintain service availability through secondary mappings. The delay during LB-PSVM is a bit higher for the reason that one node is down and the existing attack-free SIs serve additional vehicles affected by the attack which may result in the creation of queues, and subsequently adds up queuing delay. While the delay is decreased again in SRP, once the failed instances are recovered to attack-free nodes. Note that the difference in delay for LB-PSVM due to balanced processing load is negligible when compared with SP and SRP. In Fig. \ref{fig:Delay-BJ} and Fig. \ref{fig:Delay-RM}, we further evaluate the delay performance for different traffic densities using different datasets, and the performance of our framework is consistent and resilient. \par
\begin{figure}[hbt!]
	\captionsetup[subfigure]{justification=centering}
	\centering
	\begin{subfigure}{.22\textwidth}
		\centering
		\includegraphics[width=1.6in,height=1.2in]{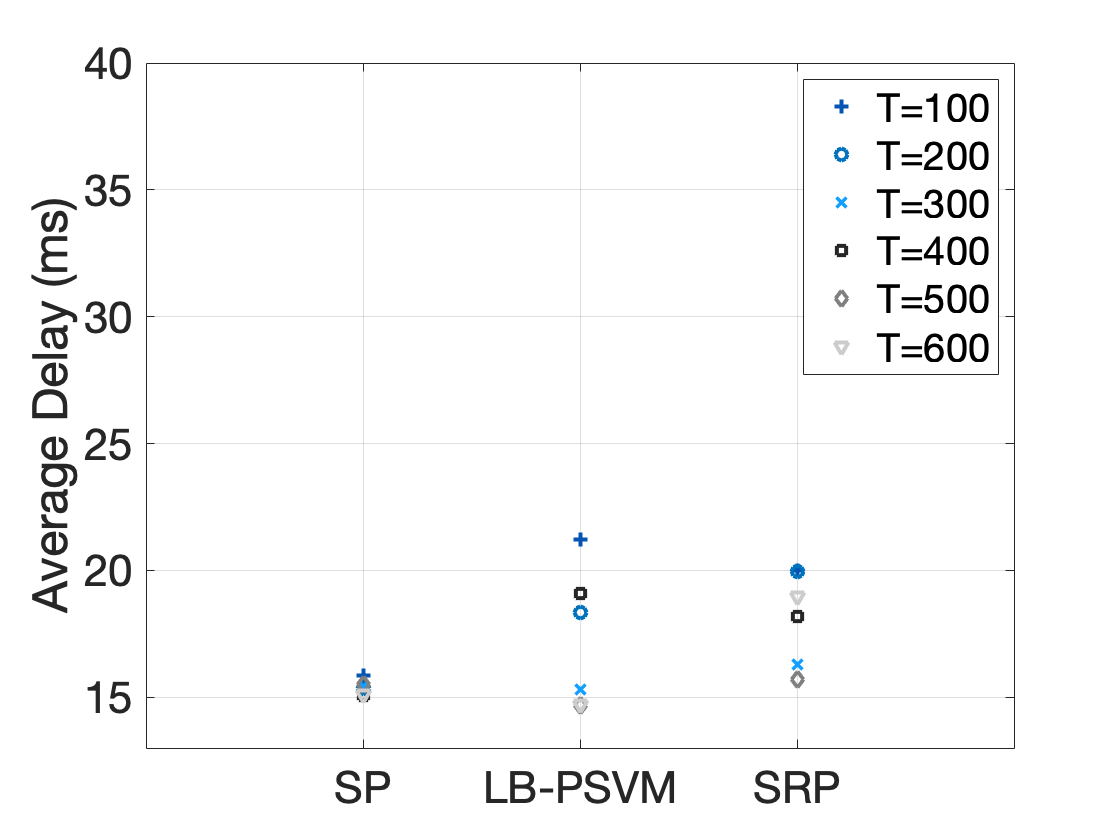}  
		\caption{San Francisco}
		\label{fig:ADelay-SF}
	\end{subfigure}
	\begin{subfigure}{.22\textwidth}
		\centering
		\includegraphics[width=1.6in,height=1.2in]{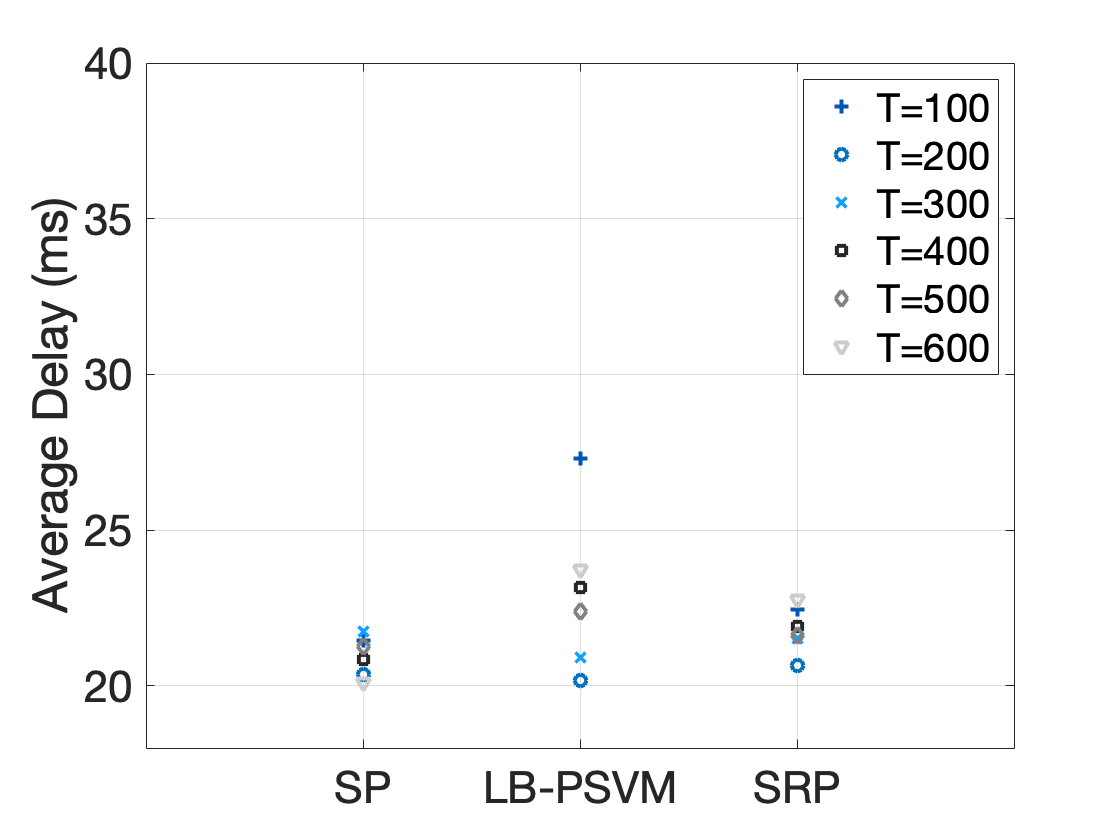}  
		\caption{Beijing}
		\label{fig:ADelay-BJ}
	\end{subfigure} 
	\begin{subfigure}{.22\textwidth}
		\centering
		\includegraphics[width=1.6in,height=1.2in]{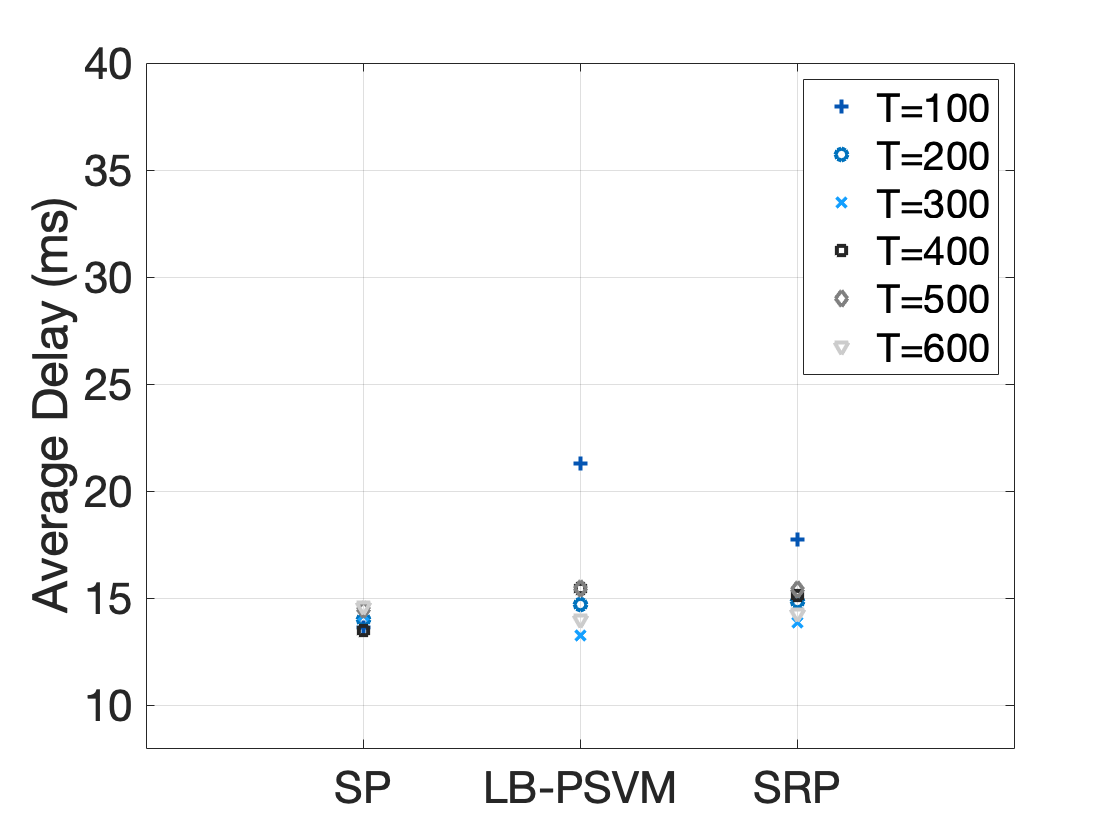}  
		\caption{Rome}
		\label{fig:ADelay-RM}
	\end{subfigure}
	\caption{Average Delay (ms)}
	\label{fig:ADelay}
\end{figure}

 
Fig. \ref{fig:ADelay} plots the average delay of all service types for different time units to observe the difference in delay for different network states. For the San Francisco dataset, the average LB-PSVM delay is $\sim$3ms-5ms higher for the 100$^{th}$, 200$^{th}$ and 400$^{th}$ time unit. On the contrary, the delay performance is more prominent for 300$^{th}$, 500$^{th}$ and 600$^{th}$ time unit where results are nearly similar to the attack-free primary mappings. When compared with recovery placements, the delay is higher for single time unit i.e. 100$^{th}$, and the difference is 1ms only. This demonstrates the effective performance of LB-PSVM during edge node failure. Similarly, for the city of Beijing and Rome, the maximum delay difference for LB-PSVM against SP and SRP, does not exceed $\sim$5ms for both. \par 
\begin{figure}[hbt!]
	\captionsetup[subfigure]{justification=centering}
	\centering
	\begin{subfigure}{.22\textwidth}
		\centering
		\includegraphics[width=1.6in,height=1.2in]{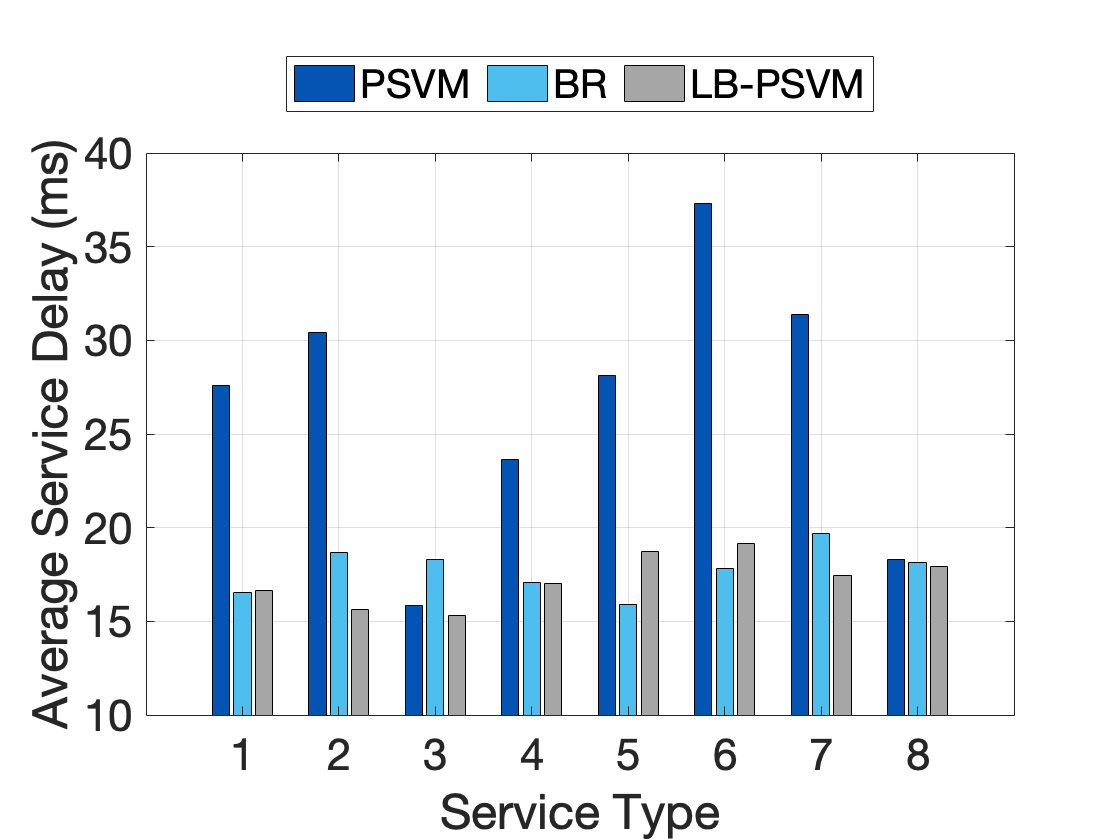}  
		\caption{San Francisco}
		\label{fig:CDelay-SF}
	\end{subfigure}
	\begin{subfigure}{.22\textwidth}
		\centering
		\includegraphics[width=1.6in,height=1.2in]{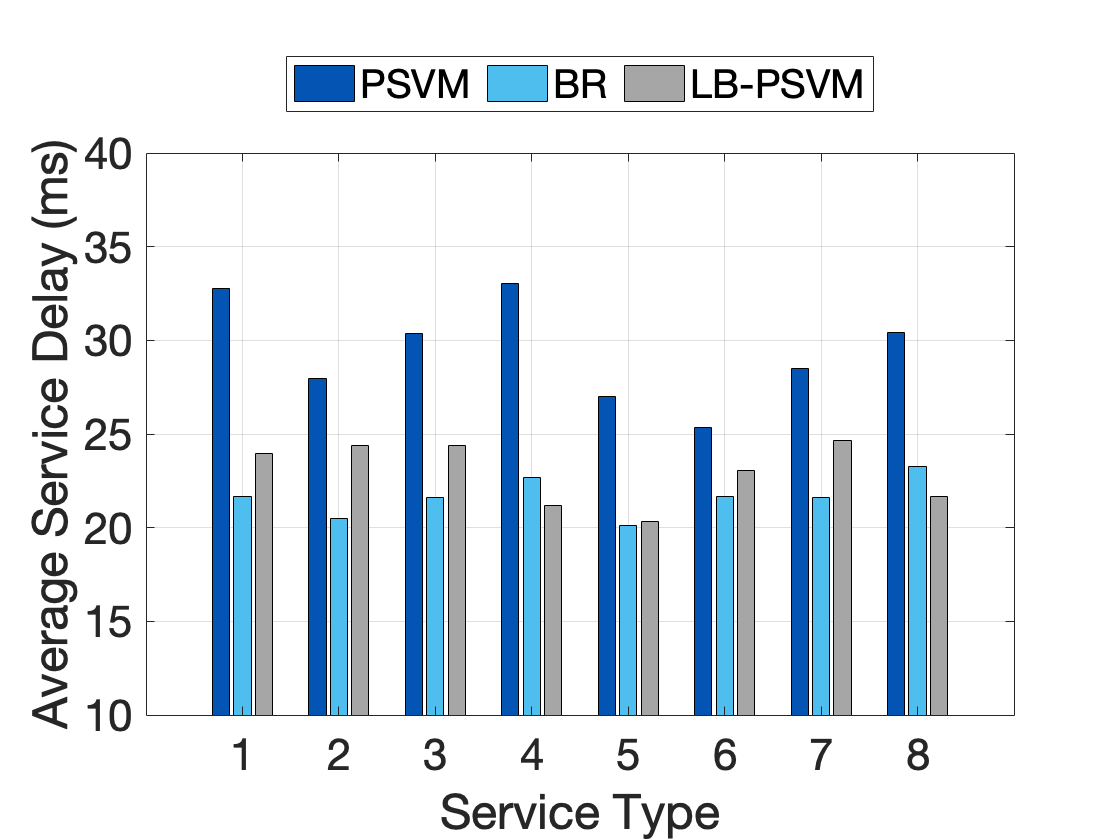}  
		\caption{Beijing}
		\label{fig:CDelay-BJ}
	\end{subfigure} 
	\begin{subfigure}{.22\textwidth}
		\centering
		\includegraphics[width=1.6in,height=1.2in]{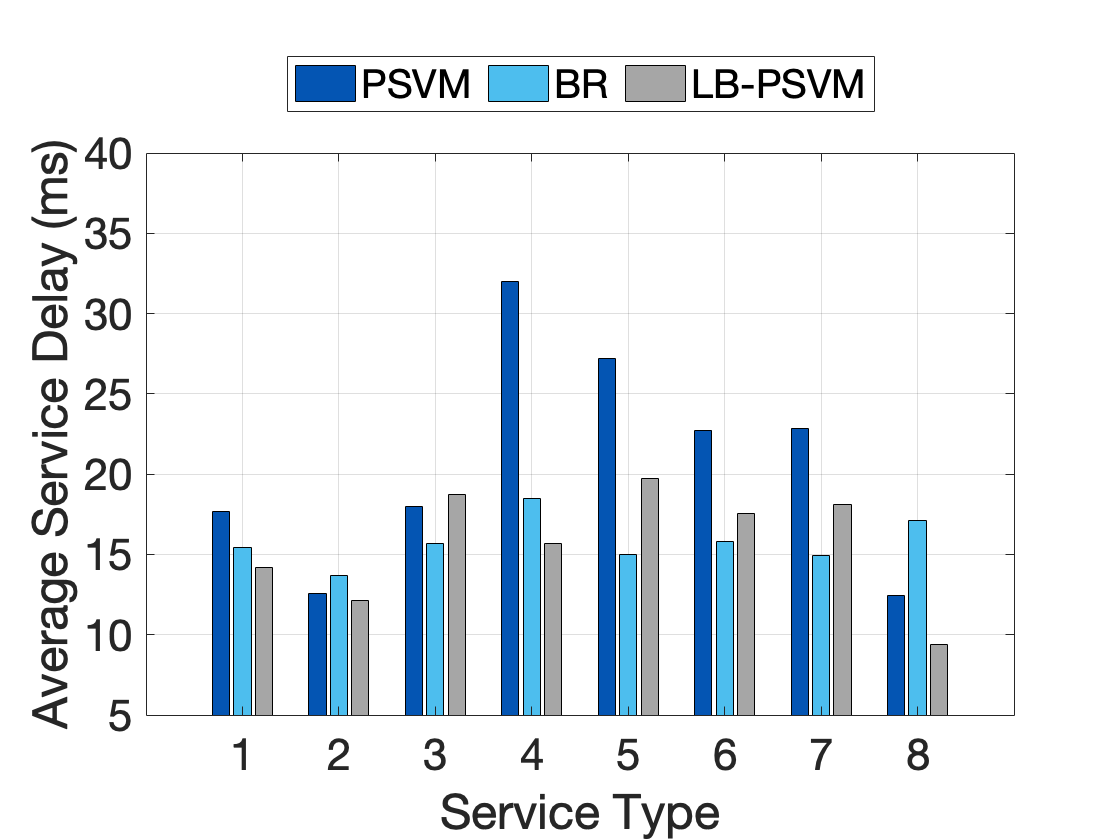}  
		\caption{Rome}
		\label{fig:CDelay-RM}
	\end{subfigure}
	\caption{Average Service Delay (ms)}
	\label{fig:CDelay}
\end{figure}

Next, we compare the LB-PSVM against the PSVM and BR. The performance metrics considered for comparison are the average service delay, edge load factor (ELF), and fairness. Fig. \ref{fig:CDelay} compares the average service delay for three different city environments. First, the delay for PSVM is the highest. However, a fairly similar performance in delay values is observed for BR and LB-PSVM but at the expense of higher edge resources usage in BR. This implies that the LB-PSVM scheme is a better choice for faster service access to the vehicles together with efficient usage of edge resources. \par
\begin{figure}[hbt!]
	\captionsetup[subfigure]{justification=centering}
	\centering
	\begin{subfigure}{.52\textwidth}
		\centering
		\includegraphics[width=1.6in,height=1.2in]{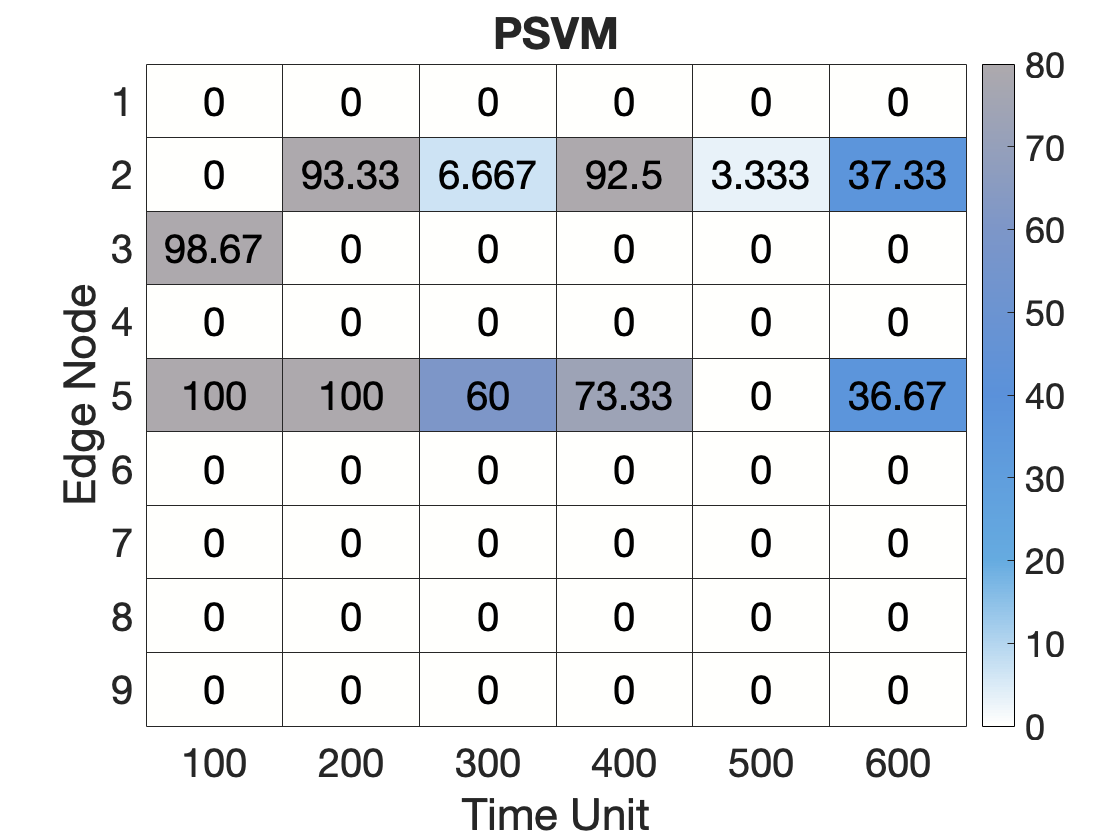}  
		\includegraphics[width=1.6in,height=1.2in]{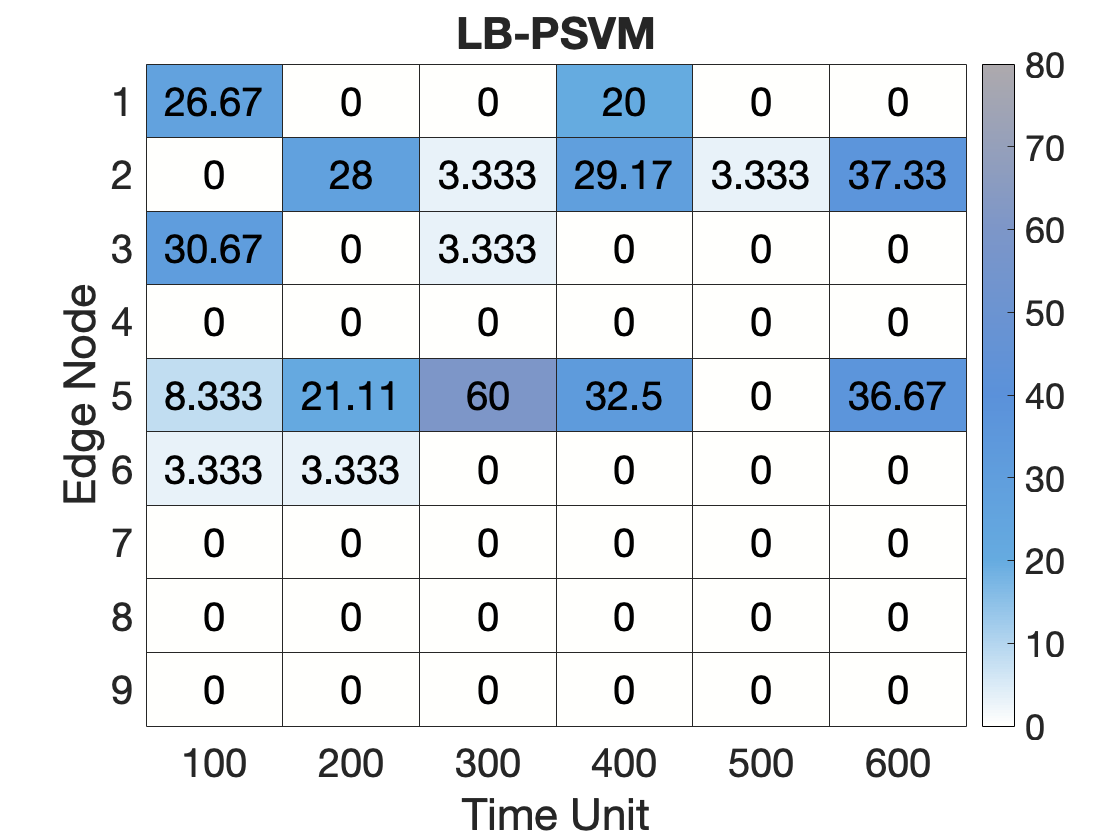}  
		\caption{San Francisco}
		\label{fig:ELF1-SF}
	\end{subfigure} \\
	\begin{subfigure}{.52\textwidth}
		\centering
		\includegraphics[width=1.6in,height=1.2in]{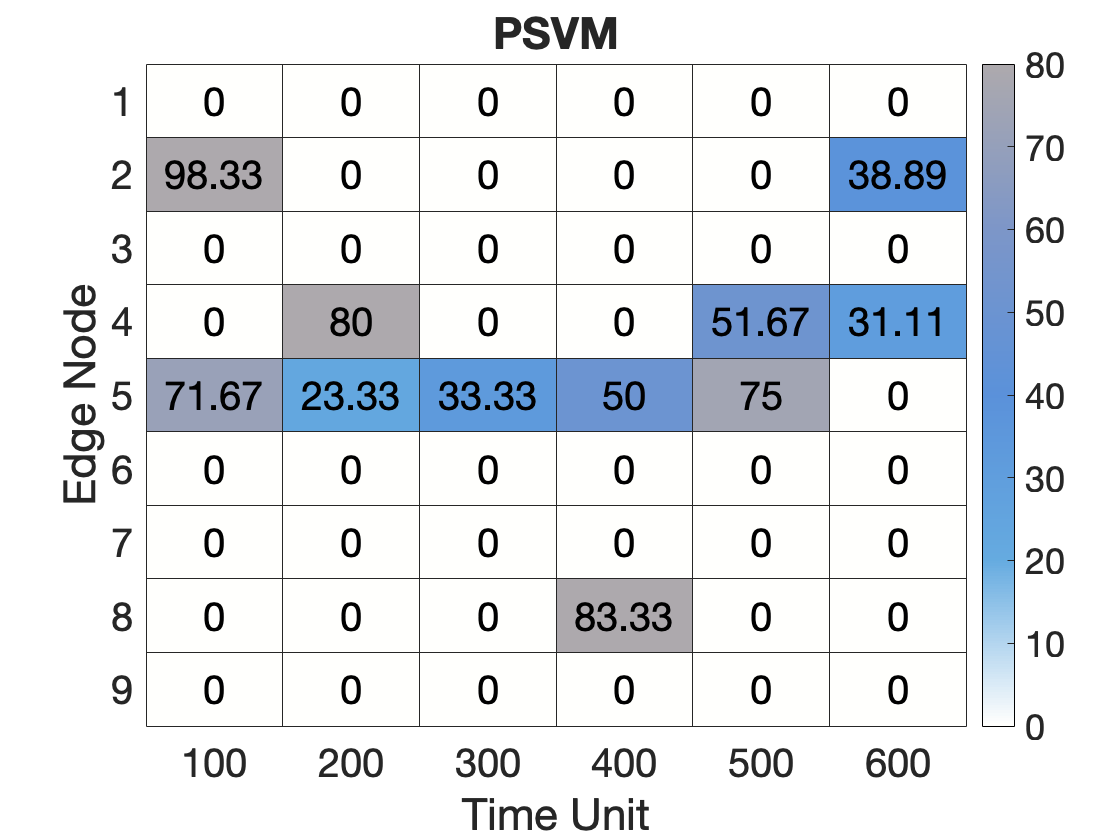}  
		\includegraphics[width=1.6in,height=1.2in]{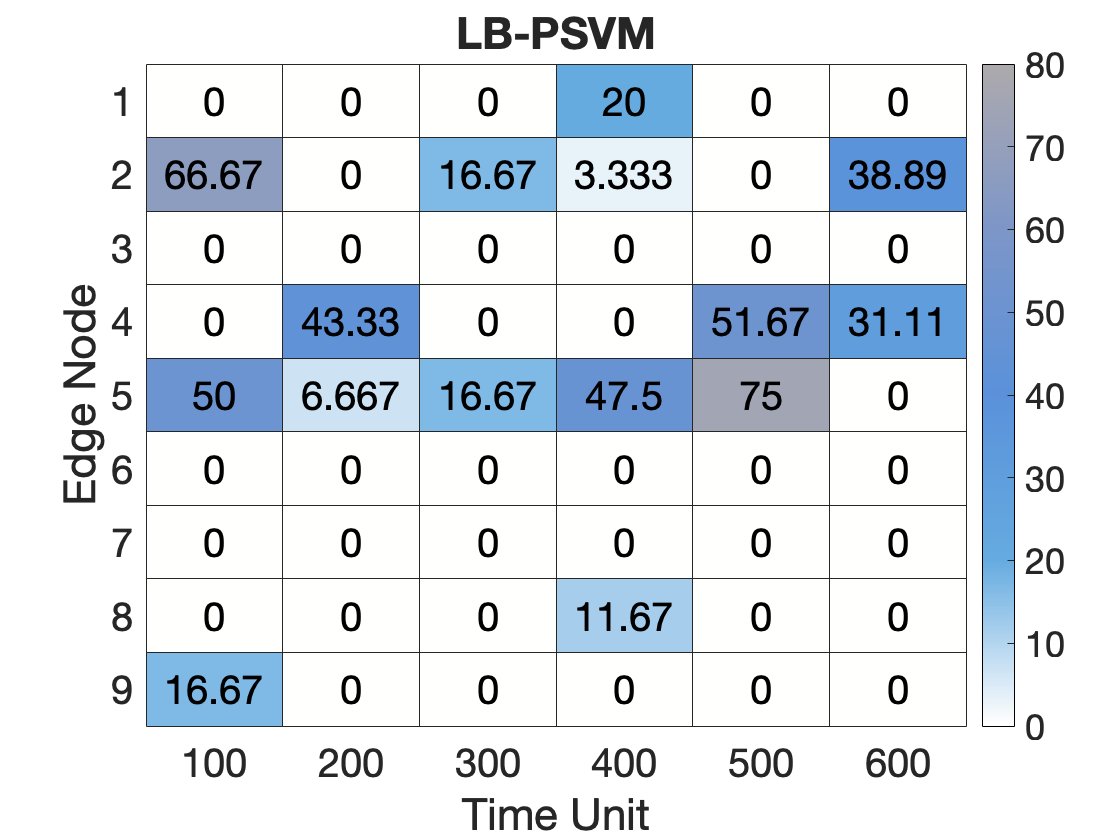}  
		\caption{Beijing}
		\label{fig:ELF1-BJ}
	\end{subfigure} 
	\begin{subfigure}{.52\textwidth}
		\centering
		\includegraphics[width=1.6in,height=1.2in]{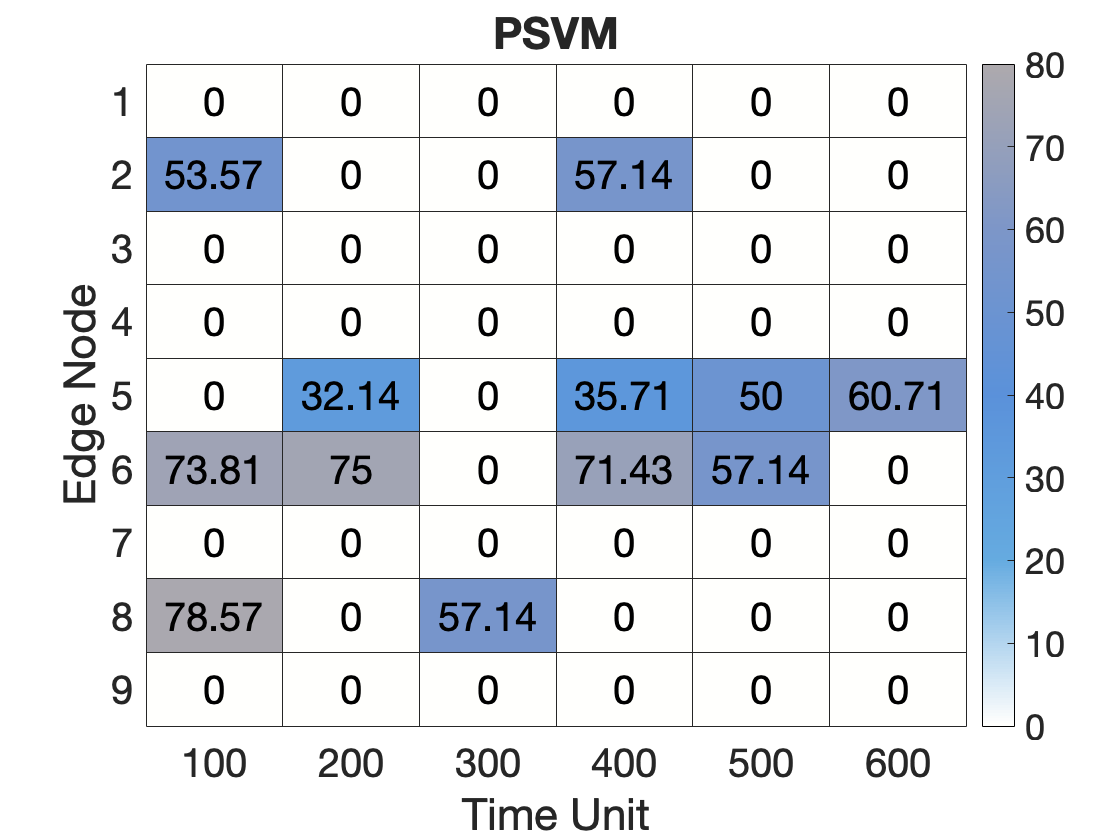} 
		\includegraphics[width=1.6in,height=1.2in]{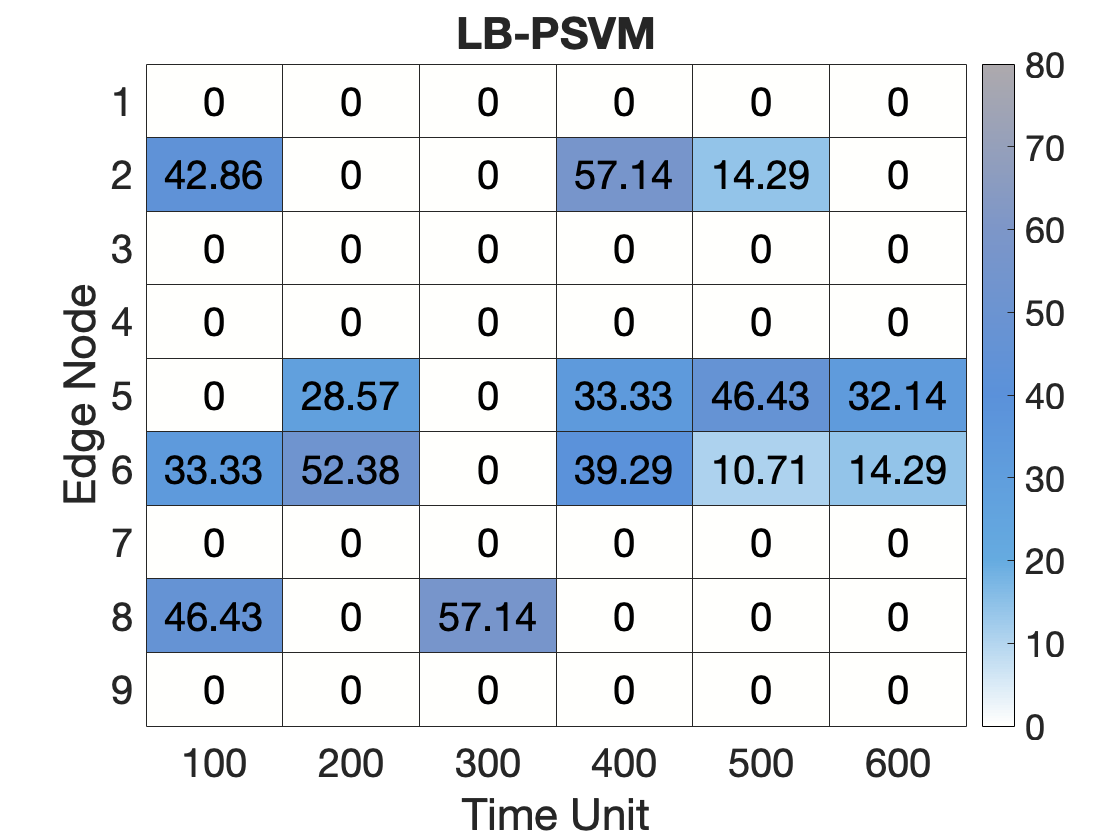}   
		\caption{Rome}
		\label{fig:ELF1-RM}
	\end{subfigure} 
	\caption{Edge Load Factor (\%)}
	\label{fig:ELF}
\end{figure}

\begin{figure}[hbt!]
	\captionsetup[subfigure]{justification=centering}
	\centering
	\begin{subfigure}{.52\textwidth}
		\centering
		\includegraphics[width=1.6in,height=1.2in]{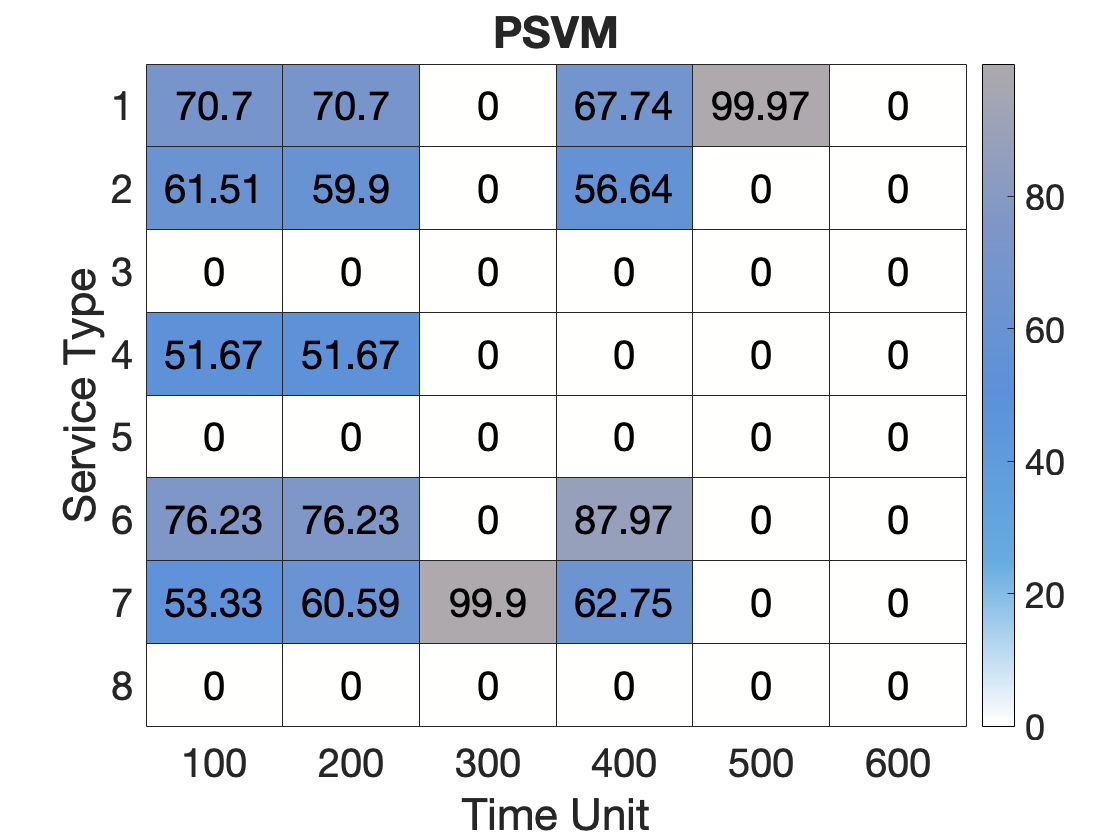}  
		\includegraphics[width=1.6in,height=1.2in]{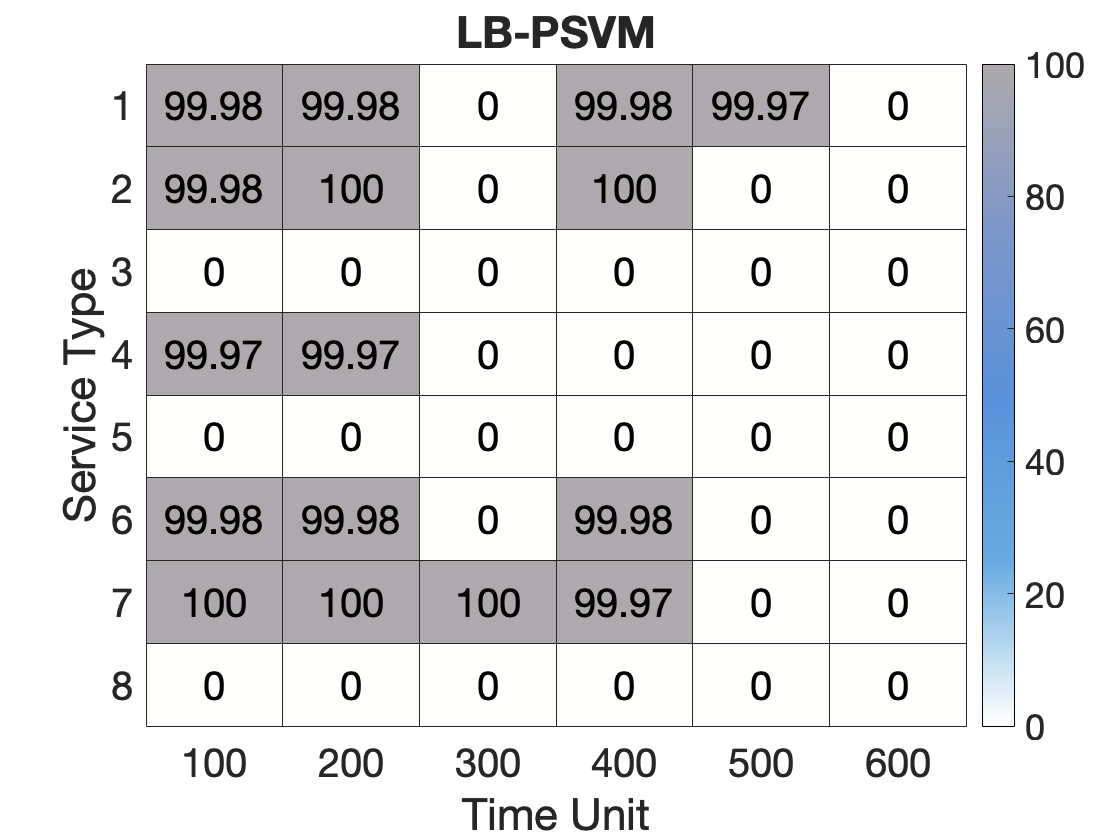}  
		\caption{San Francisco}
		\label{fig:F1-SF}
	\end{subfigure}
	\begin{subfigure}{.52\textwidth}
		\centering
		\includegraphics[width=1.6in,height=1.2in]{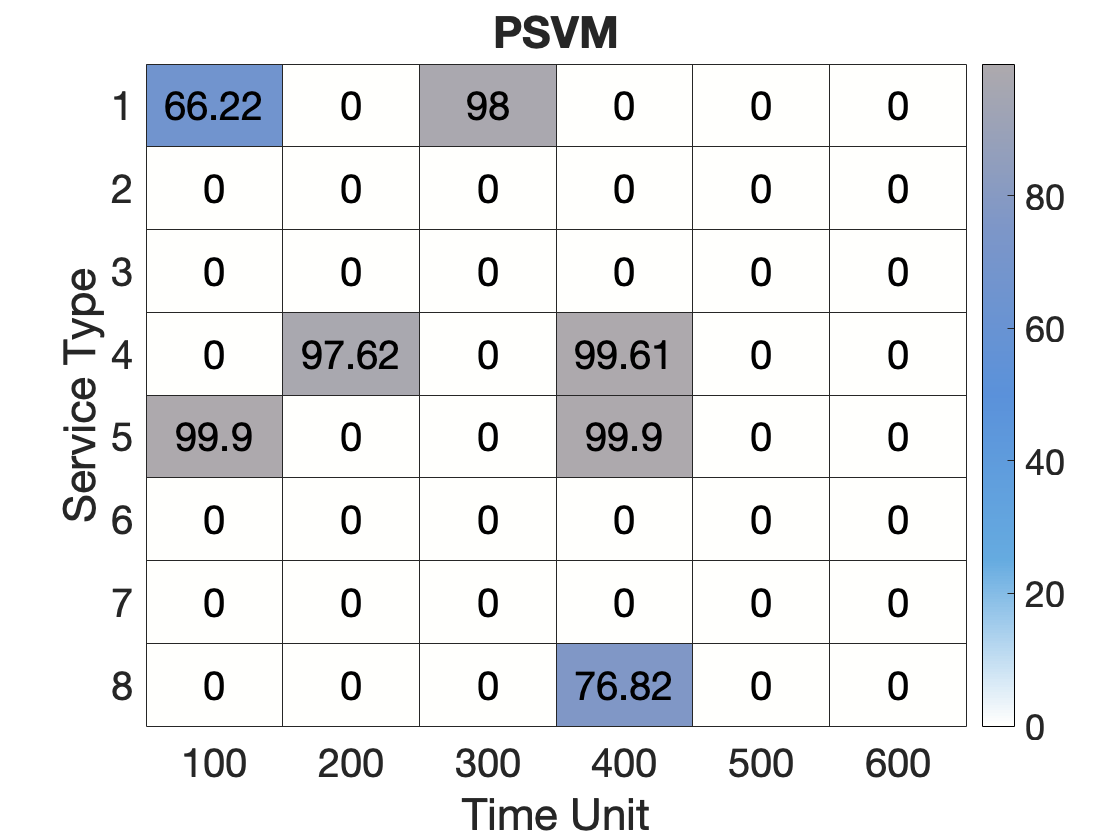}  
		\includegraphics[width=1.6in,height=1.2in]{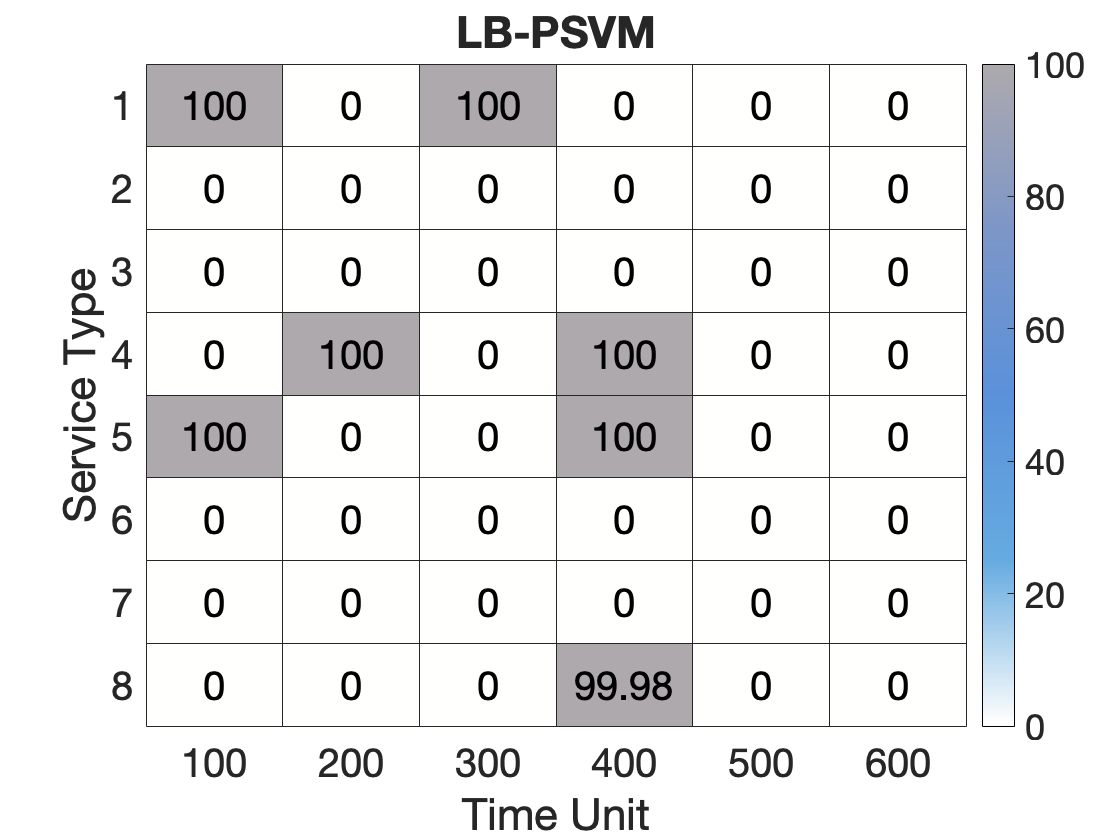}  
		\caption{Beijing}
		\label{fig:F1-BJ}
	\end{subfigure} 
	\begin{subfigure}{.52\textwidth}
		\centering
		\includegraphics[width=1.6in,height=1.2in]{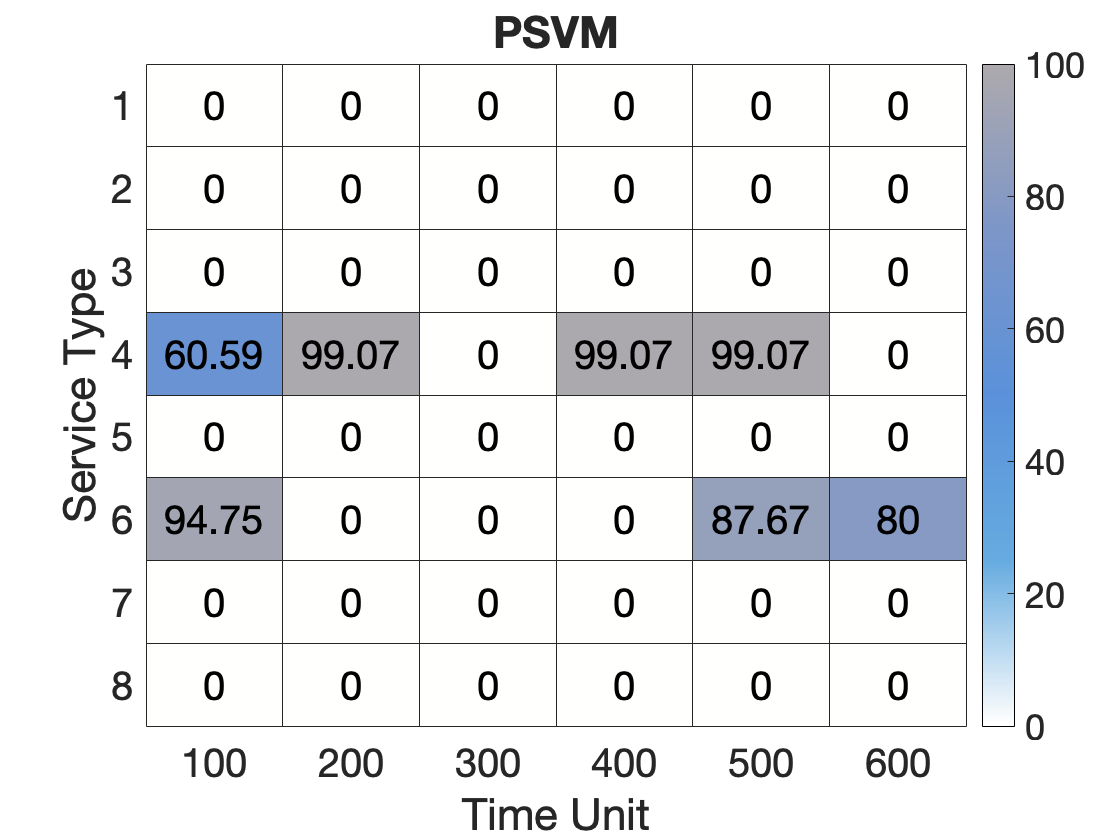} 
		\includegraphics[width=1.6in,height=1.2in]{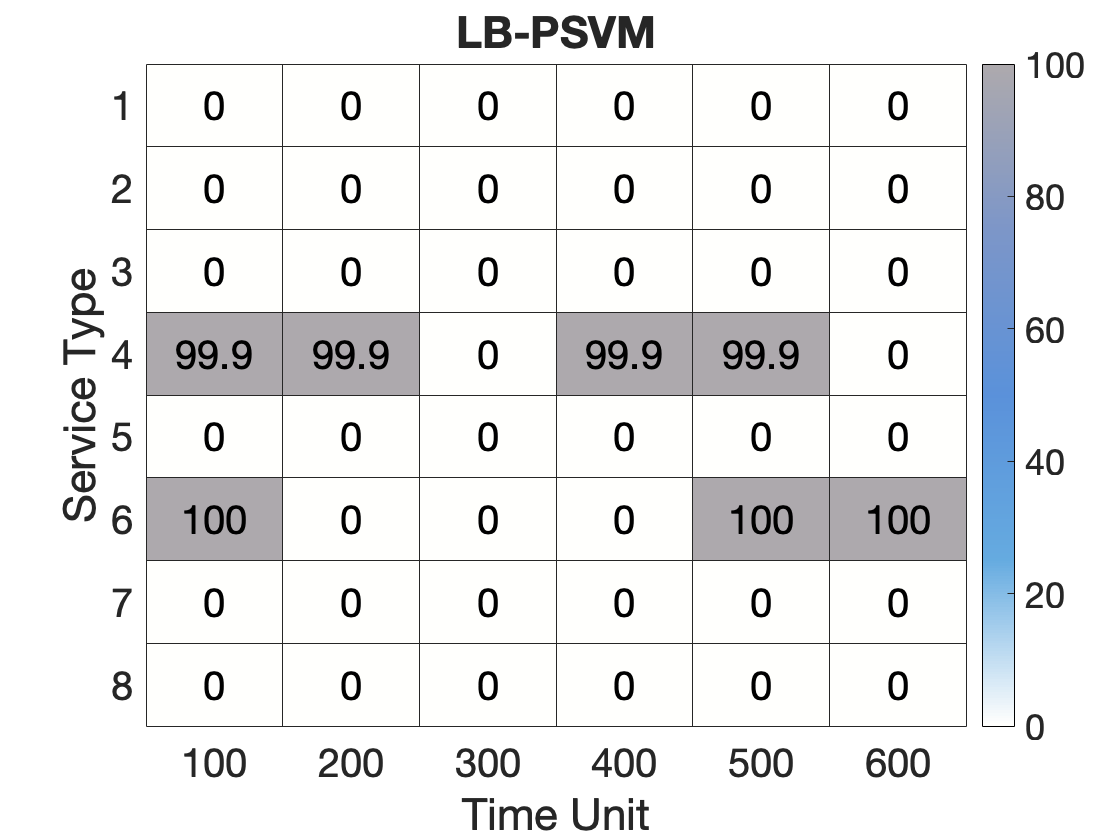}   
		\caption{Rome}
		\label{fig:F1-RM}
	\end{subfigure} \\
	\caption{Fairness}
	\label{fig:F}
\end{figure}

As the BR approach does not consider the selection of edge nodes for secondary mappings and rather uses pre-reserved SI, we cannot compare our PSVM and LB-PSVM with BR in terms of ELF and fairness. Fig. \ref{fig:ELF} and Fig. \ref{fig:F} depict the percentage of ELF and fairness, respectively. The load factor for each SI is calculated as the ratio of additional processing load to the SI's available processing capacity. The ELF is the average of the load factor for all SIs on an edge node. The fairness represents how fairly the vehicles are mapped across different attack-free edge nodes for secondary mappings while taking the primary load of edge into account. We use Jain's index as a fairness measure in this work. \par 
Fig. \ref{fig:ELF} plots the heat map of ELF for each node against different time units to indicate the percentage variation of load on different edge nodes and in different datasets. In the PSVM, most of the attacked-affected vehicles are clustered over the limited number of edge nodes resulting in saturation with an average load greater than 60\% most of the time. On the contrary, with our proposed LB-PSVM model, we consider the fair usage of edge processing capacities in serving vehicle requests across multiple edge nodes to avoid the state of congestion on a single edge node. The effectiveness of LB-PSVM is demonstrated with the spread of requests among different edge nodes. As an example, in the city of San Franciso, at 100$^{th}$ time unit, the same number of attack-influenced vehicles (for different service types) are mapped to two edge nodes with a maximum load of 100\% in PSVM. On the contrary, with LB-PSVM, the same demand is fulfilled in a fairly distributed way among 4 different edge nodes limiting the maximum load to 30\%. \par 
To study the average additional load shifted to the working SIs from the failed SIs, we plot the average ELF values in Table \ref{tab:averageedgeload}. The results show that PSVM exhibits imbalanced and overloading behaviour compared to the proposed LB-PSVM. We note that inefficient use of available processing capacities limits the heavily-loaded nodes from accommodating future service requests which will be forced to be accessed from farther edge nodes resulting in higher delays and poor performance. \par

\begin{table}[htbp]
	\centering
	\caption{Average Edge Load Factor (\%)}
	\scriptsize
	\tabcolsep=0.09cm
	\scalebox{0.8}{
		\begin{tabular}{clcccccc}
			\toprule
			\multicolumn{2}{c}{} & \textbf{T=100} & \textbf{T=200} & \textbf{T=300} & \textbf{T=400} & \textbf{T=500} & \textbf{T=600} \\
			\midrule
			\multirow{2}[4]{*}{\textbf{San Francisco}} & \textbf{PSVM} & 99.33 &  96.66  & 33.33  & 82.91  &  3.33   & 37.00 \\
			\cmidrule{2-8}          & \textbf{LB-PSVM} & 17.25 & 17.48 & 22.22 & 27.22 & 3.33 &  37.00 \\
			\midrule
			\multirow{2}[4]{*}{\textbf{Beijing}} & \textbf{PSVM} & 85.00 &  51.66 &  33.33 & 66.66 & 63.33 & 35.00\\
			\cmidrule{2-8}          & \textbf{LB-PSVM} & 44.44 & 25.00 & 16.66 &  20.62 &  63.33 & 35.00 \\
			\midrule
			\multirow{2}[4]{*}{\textbf{Rome}} & \textbf{PSVM} & 68.65 & 53.57 & 57.14 & 54.76 & 53.57 & 60.71 \\
			\cmidrule{2-8}          & \textbf{LB-PSVM} & 40.87 & 40.47 &  57.14 &  43.25 & 23.80 &  23.21 \\
			\bottomrule
		\end{tabular}%
	}
	\label{tab:averageedgeload}%
\end{table}%

Fig. \ref{fig:F} depicts the level of load balancing in terms of fairness among different edge nodes while mapping attack-affected vehicles. It validates that the fairness in allocating load over different edge nodes in LB-PSVM is always greater than 90\%. However, in PSVM, it is quite low, and it is as low as 50\%. It shows that LB-PSVM performs better in this regard since it takes into consideration both delay and balanced use of edge processing capacities while determining the secondary mappings. 

\section{Conclusion}
In this paper, we developed an optimization formulation for secondary V2E mapping for attack-resilient service placement in an edge-enabled IoV network. Our solution is used in a DRL-based framework considering variations in service demands and vehicle mobility. In our proposed secondary mapping model (i.e. LB-PSVM), the affected service requests hosted on an attacked server are distributed among the attack-free working edge nodes until recovery takes place. We formulated an optimization mapping problem with the objective to i) provide faster and disruption-free service access to the vehicles upon an attack and ii) maintain balanced load among edge servers. We carried out extensive performance study using real-world datasets and evaluated our approach by comparing it against the PSVM model and BR approach and showed that the proposed LB-PSVM performs better in reducing the delay while achieving better load balancing at the edge nodes. 

\balance

\bibliographystyle{IEEEtran}
\bibliography{IEEEabrv,References}

\end{document}